\documentclass[aps,prd,preprintnumbers,showpacs,showkeys,nofootinbib,
superscriptaddress,fleqn,floatfix,tightenlines,10pt]{revtex4-1}
\usepackage{amsmath,amsfonts,amssymb,amscd,amsxtra,amsthm}
\usepackage{graphicx}  
\usepackage{epstopdf}
\usepackage{dcolumn}  
\usepackage{bm}          
\usepackage{slashed}
\usepackage{cancel}
\usepackage{float}
\usepackage{mathtools}
\usepackage{amsbsy}
\usepackage{amstext}

\usepackage[utf8]{inputenc}
\usepackage{booktabs}
\usepackage[normalem]{ulem} 
\usepackage[dvipsnames]{xcolor} 
\usepackage{tabularx}
\usepackage{enumitem}
\usepackage{array}
\usepackage{slashed}
\usepackage{tikz}
\usepackage{float}
\usepackage{multirow}

\newcommand{\Psfig}[2]{\includegraphics[width=#1]{#2}}
\newcommand{\PsfigII}[2]{\includegraphics[scale=#1]{#2}}

\def\mev{\text{ MeV}}
\def\gev{\text{ GeV}}

\def\microb{~\mu \text{b}}
\def\millib{\text{ mb}}

\begin{document}
\preprint{INHA-NTG-12/2019}
\title{Feasibility study of the $K^{+} d \to K^{0} p p$ reaction
  for the ``$\Theta ^{+}$'' pentaquark}
\author{Takayasu~Sekihara}
\email{sekihara@post.j-parc.jp}
\affiliation{Advanced Science Research Center, Japan Atomic
  Energy Agency, Shirakata, Tokai, Ibaraki, 319-1195, Japan}
\affiliation{Research Center for Nuclear Physics (RCNP), Osaka University,
  Ibaraki, Osaka, 567-0047, Japan}
\affiliation{RIKEN, Wako, Saitama, 351-0198, Japan}

\author{Hyun-Chul Kim}
\email{hchkim@inha.ac.kr}
\affiliation{Department of Physics, Inha University, Incheon 22212,
  Republic of Korea}
\affiliation{Advanced Science Research Center, Japan Atomic
  Energy Agency, Shirakata, Tokai, Ibaraki, 319-1195, Japan}
\affiliation{School of Physics, Korea Institute for Advanced Study (KIAS),
    Seoul 02455, Republic of Korea}
  \author{Atsushi Hosaka}
  \email{hosaka@rcnp.osaka-u.ac.jp}
  \affiliation{Research Center for Nuclear Physics (RCNP), Osaka University,
    Ibaraki, Osaka, 567-0047, Japan}
\affiliation{Advanced Science Research Center, Japan Atomic
  Energy Agency, Shirakata, Tokai, Ibaraki, 319-1195, Japan}
\date{April, 2020}
\begin{abstract}%
  We investigate theoretically the $K^{0} p$ invariant mass spectrum
  of the $K^{+} d \to K^{0} p p$ reaction and scrutinize how the
  signal of the ``$\Theta ^{+}$'' pentaquark, if it exists, emerges in
  the $K^{0} p$ spectrum.  The most prominent advantage of this
  reaction is that we can clearly judge whether the ``$\Theta ^{+}$''
  exists or not as a direct-formation production without significant
  backgrounds, in contrast to other reactions such as photoproduction
  and $\pi$-induced productions.  We show that while the impulse or
  single-step scattering process can cover the ``$\Theta ^{+}$''
  energy region with an initial kaon momentum $k_{\rm lab} \approx
  0.40 \gev / c$ in the laboratory frame, the contributions from
  double-step processes may have a potential possibility to reach the
  ``$\Theta ^{+}$'' energy region with a higher kaon momentum $k_{\rm
    lab} \sim 1 \gev / c$.  Assuming that the full decay width of the
  ``$\Theta ^{+}$'' is around $0.5 \mev$, we predict that the
  magnitude of the peak corresponding to the ``$\Theta^+$'' is around
  a few hundred $\mu \text{b}$ to $1 \millib$ with the momentum of the
  kaon beam $k_{\rm lab} \approx 0.40 \gev / c$ while it is around
  $\lesssim 1 \microb$ with $k_{\rm lab} \approx 0.85 \gev / c$.
  Thus, the  ``$\Theta^+$'' peak is more likely to be seen at
  $k_{\mathrm{lab}}   \approx 0.40 \gev / c$ than at $k_{\mathrm{lab}}
  \approx 0.85 \gev /   c$.
\end{abstract}

\maketitle

\section{Introduction}
The physics of pentaquarks, which are the baryons consisting of four
valence quarks and one anti-quark, has been renewed very recently, as
the LHCb Collaboration announced the new findings of three heavy
pentaquarks, $P_c$'s~\cite{Aaij:2015tga, Aaij:2016phn, Aaij:2016ymb,
  Aaij:2019vzc}. The LHCb Collaboration also found the five excited
$\Omega_c$'s in the channel of the $\Xi_c^+ K^-$ invariant
mass~\cite{Aaij:2017nav}. The four of them were confirmed by the Belle
Collaboration~\cite{Yelton:2017qxg}.  Since these newly found excited
$\Omega_c$'s have very small decay widths, several theoretical
works have suggested that at least some of them may be identified as
the singly heavy pentaquarks~\cite{Kim:2017jpx, Kim:2017khv,
  An:2017lwg, Wang:2018alb}.  On the other hand, the discussion of the
light pentaquarks became dormant, which was once triggered by the
theoretical prediction~\cite{Diakonov:1997mm} and the first
measurement of the ``$\Theta^+$''~\cite{Nakano:2003qx}, since the null
results of the ``$\Theta^+$'' baryon reported by the CLAS
Collaboration~\cite{Battaglieri:2005er, McKinnon:2006zv,
  DeVita:2006aaq}. Moreover, both the KEK-PS E533
Collaboration~\cite{Miwa:2006if} and the J-PARC E19
Collaboration~\cite{Shirotori:2012ka, Moritsu:2014bht} searched for
the ``$\Theta^+$'' using the pion beam but found no significant peak
corresponding to the ``$\Theta^+$'' pentaquark. The Belle
Collaboration looked for isospin partners of the ``$\Theta^+$'' in the
first observed process $\gamma \gamma \to p \bar{p} K^+ K^-$ but again
has no significant evidence for them~\cite{Shen:2016csu}. All these
negative results make the existence of the ``$\Theta^+$'' rather
skeptical, so that both experimental and theoretical investigations on 
the ``$\Theta^+$'' ebbed away.

In the meanwhile, the LEPS Collaboration and DIANA Collaboration
continued to report the evidence for the existence of the
``$\Theta^+$''~\cite{Nakano:2008ee, Niiyama:2013dya, Barmin:2013lva,
  Barmin:2015cta}. Some years ago, Amaryan et al.\ analyzed the data
from the CLAS Collaboration~\cite{Amaryan:2011qc}, using the
interference method with $\phi$-meson photoproduction. They found the
peak around $\sim 1.54$ GeV, which corresponds to the
``$\Theta^+$''. The statistical significance of this peak was
$5.3\,\sigma$~\cite{Amaryan:2011qc}.  In Ref.~\cite{Asratyan:2016qfs},
the SELEX data on hadro-nucleus collisions at Fermilab were analyzed
in searching for formation of the ``$\Theta^+$''. A narrow enhancement
near 1539 MeV was observed in the mass spectrum of the $pK_S^0$ system
emitted at small $x_F$ from hadron collisions with copper nuclei,
where $x_F$ denotes the Feynman variable defined as the ratio of the
momentum $p_L^*/p_{\mathrm{max}}^*$ (For details see
Ref.~\cite{Asratyan:2016qfs}). However, the results from
Ref.~\cite{Asratyan:2016qfs} show definite dependence on the
kinematics.

After the LHCb Collaboration reported the existence of the heavy
pentaquarks, interest in light pentaquarks seems to be renewed. 
For example, the existence of a narrow nucleon resonance $N^*(1685)$
has been announced by a series of experiments in $\eta$
photoproduction off the
quasi-neutron~\cite{Kuznetsov:2006kt,Miyahara:2007zz,
  Kuznetsov:2008hj, Jaegle:2008ux, Jaegle:2011sw,Werthmuller:2013rba,
  Witthauer:2013tkm, Werthmuller:2014thb,McNicoll:2010qk,
  Witthauer:2017get}. More recently, a similar narrow peak was observed
in the $\gamma p \to p\pi^0 \eta$
reaction~\cite{Metag:2019bay}. Though the identification of this
narrow resonance is still under debate, one possible interpretation is
that it can be regarded as a pentaquark nucleon, which is a member of
the baryon antidecuplet~\cite{Polyakov:2003dx, Kim:2005gz,Yang:2018gju}. 
     
Based on previous experimental studies on the ``$\Theta^+$'', 
we could draw
at least one conclusion: the ``$\Theta^+$'' is most unlikely to
exist. Even though it might exist, it is elusive to observe it. 
However, we want to mention that almost all previous experiments
have utilized indirect methods such as the photon and pion beams,
which suffer from large backgrounds~\cite{MartinezTorres:2010zzb,Torres:2010jh}. 
Moreover, we know that the
``$\Theta^+$'', if it exists, decays into $K^0 p$ or $K^+ 
n$. In particular, the $K^+ n$ channel may be the 
most probable one to search for the ``$\Theta^+$''. In fact, the DIANA
Collaboration used the low-energy $K^+ \mathrm{Xe}$ reaction in the
xenon bubble chamber~\cite{Barmin:2003vv, Barmin:2006we,
  Barmin:2009cz, Barmin:2013lva, Barmin:2015cta}, though there is also
a theoretical criticism on the DIANA results in
2003~\cite{Sibirtsev:2004cf}. Nevertheless,     
the $K^+$ beam may provide an ultimate smoking gun whether the
``$\Theta^+$'' exists or not, since it will create the ``$\Theta^+$'' by
direct formation and will be seen in the differential and total cross
sections, if it exists. 
There is also a discussion on $K^+N$ reactions in a nucleus~\cite{Gal:2005cz}.
Thus, measuring the $K^+ d \to K^0
pp$ reaction is the 
simplest and
final experiment to put a period to the
existence of the ``$\Theta^+$'' pentaquark. This process, compared with
other reactions such as photoproduction and $\pi$-induced productions,
is not hampered by significant backgrounds. This means that the
experiment of the $K^+d \to K^0 pp$ will clearly judge the existence
of the ``$\Theta^+$''.        

The $K^+ d \to K^0 pp$ reaction was already investigated
theoretically~\cite{Sibirtsev:2004bg} with the width of the
``$\Theta^+$'' being assumed to be $1$--$20 \mev$. 
Simulations were also performed for experiments proposed at J-PARC~\cite{E949:2014xx,P09-LoI:2014xx}.  
In Ref.~\cite{Sibirtsev:2004bg}
Sibirtsev et al.\ considered the single-step process or the impulse
scattering process in which the proton in the deuteron was regarded as
a spectator and the neutron interacts with the $K^+$ to produce the
proton and the neutral kaon. When the $K^+$ momentum lies in the range
of $0.47$--$0.64 \gev / c$, the peak corresponding to the
``$\Theta^+$'' was seen in the $K^0 p$ invariant mass spectra. In the
vicinity of $0.47 \gev / c$, the ``$\Theta^+$'' peak was also shown in
the total cross sections. In the present work, we include both the
single-step and double-step processes and scrutinize the feasibility
of the $K^+ d \to K^0 pp$ reaction in observing the ``$\Theta^+$''
pentaquark. 
The use of the double step process was also proposed in Ref.~\cite{JPARC-LoI:2018xx}. 
In the double-step processes, a kaon is exchanged in the
course of the interaction between the proton and the neutron. We will
show that in the present work the single-step process can cover the
energy region corresponding to the ``$\Theta^+$'' peak with an initial
kaon momentum $k_{\mathrm{lab}}\approx 0.4 \gev /c$ in the laboratory
(Lab)
frame while the double-step processes provide a potential possibility
to reach the ``$\Theta^+$'' energy region with a higher kaon momentum 
$k_{\mathrm{lab}}\approx 1 \gev /c$. In the present work, thus, we
will carefully investigate the $K^+d\to K^0 pp$ reaction in the
context of a possible existence of the ``$\Theta^+$'', considering
both the single- and double-step processes.

This paper is organized as follows.  In Sec.~\ref{sec:2}, we formulate
the cross section of the $K^{+} d \to K^{0} p p$ reaction.  The $K N
\to K N$ scattering amplitude is also shown in this section.  In
Sec.~\ref{sec:3}, we give numerical results on the cross section of
the $K^{+} d \to K^{0} p p$ reaction and investigate strength of a
peak signal in the $K^{0} p$ spectrum, which will provide a good
guideline to conclude whether the ``$\Theta ^{+}$'' exists or not. 
Section~\ref{sec:4} is devoted to the summary of this study.

\section{Formulation}
\label{sec:2}

\subsection{Cross section of the $K^{+} d \to K^{0} p p$ reaction}
\label{sec:2-1}

First of all, we formulate the cross section of the $K^{+} d \to K^{0}
p p$ reaction.  Since we are interested in the $K^{0} p$ invariant
mass spectrum, in which we search for the ``$\Theta ^{+}$'' signal, it
is convenient to calculate the differential cross section as a
function of the $K^{0} p$ invariant mass together with the scattering
angle for the other proton.  In this respect, we can express the
differential cross section of this reaction as~\cite{Jido:2009jf,
  Jido:2010rx, Jido:2012cy, YamagataSekihara:2012yv},
\begin{equation}
  \frac{d^{2} \sigma}{d M_{K^{0} p} d \cos \theta _{2}^{\prime}}
  = \frac{m_{d} m_{p}^{2}}{64 \pi ^{4} k_{\rm cm} W^{2}}
  p_{2}^{\prime} p_{K}^{\ast} \int d \Omega _{K}^{\ast}
  | \mathcal{T} | ^{2} .
  \label{eq:ds}
\end{equation}
Before we explain Eq.~\eqref{eq:ds}, let us distinguish the two
protons in the final state. We will call the proton that is involved
in producing the ``$\Theta^+$'' together with $K^0$ as the ``first''
proton, whereas the other one is called as the ``second'' proton.
$M_{K^{0} p}$ in Eq.~\eqref{eq:ds} denotes the invariant mass of the
$K^{0}$ and  ``first'' $p$, $\theta _{2}^{\prime}$ stands for the
scattering angle for the ``second'' proton in the center-of-mass (CM)
frame of the $K^+ d$ system, and $\Omega _{K}^{\ast}$ represents the 
solid angle for the $K^{0}$ in the rest frame of the $K^{0}$ and first
$p$.  $W$ is the CM energy of the $K^+ d$ system, and $m_{d}$ and
$m_{p}$ correspond to the masses of deuteron and proton, respectively.
The prefactor of the cross section contains the following momenta: the
initial kaon momentum $k_{\rm cm}$ and the final second proton
momentum $p_{2}^{\prime}$ are evaluated within the CM
frame, while the final kaon momentum $p_{K}^{\ast}$ is obtained in
the rest frame of the $K^{0}$ and first $p$.  They are calculated as
\begin{equation}
  k_{\rm cm} = \frac{\lambda ^{1/2} ( W^{2} , \, m_{K^{+}}^{2}, \,
    m_{d}^{2})} {2 W} ,
  \quad
  p_{2}^{\prime}
  = \frac{\lambda ^{1/2} ( W^{2} , \, M_{K^{0} p}^{2} , \, m_{p}^{2})}
  {2 W} ,
  \quad
  p_{K}^{\ast} = \frac{\lambda ^{1/2} 
      ( M_{K^{0} p}^{2} ,\, m_{K^{0}}^{2},\, m_{p}^{2})}{2 M_{K^{0} p}} ,
\end{equation}
where $\lambda ( x , \, y , \, z ) \equiv x^{2} + y^{2} + z^{2} - 2 x
y - 2 y z - 2 z x$.  $| \mathcal{T} |^{2}$ denotes the squared
scattering amplitude.

When we calculate the $K^{0} p$ invariant mass spectrum or total cross
section, we need a factor $1/2$ to avoid the double counting of the
two protons in the final state:
\begin{equation}
  \frac{d \sigma}{d M_{K^{0} p}} = \frac{1}{2} \int \cos \theta _{2}^{\prime}
  \frac{d^{2} \sigma}{d M_{K^{0} p} d \cos \theta _{2}^{\prime}} ,
  \quad
  \sigma = \frac{1}{2} \int d M_{K^{0} p} \int \cos \theta _{2}^{\prime}
  \frac{d^{2} \sigma}{d M_{K^{0} p} d \cos \theta _{2}^{\prime}} .
\end{equation}

In addition to the differential cross section $d^{2} \sigma / d
  M_{K^{0} p} d \cos \theta _{2}^{\prime}$, we may consider $d^{2}
  \sigma / d M_{K^{0} p (1)} d M_{K^{0} p (2)}$ as a Dalitz plot of
  the $K^{+} d \to K^{0} p p$ reaction, where $M_{K^{0} p (1)}$
  ($M_{K^{0} p (2)}$) is the invariant mass of the $K^{0}$ and
  ``frist'' (``second'') proton.  We can calculate this by the formula
  \begin{equation}
    \frac{d^{2} \sigma}{d M_{K^{0} p (1)} d M_{K^{0} p (2)}}
    = \frac{m_{d} m_{p}^{2}}{128 \pi ^{4} k_{\rm cm} W^{3}}
    M_{K^{0} p (1)} M_{K^{0} p (2)} \int d \cos \theta _{2}^{\prime}
    \int d \phi _{K}^{\ast} | \mathcal{T} | ^{2} ,
  \end{equation}
  where $\phi _{K}^{\ast}$ is the azimuthal angle for the $K^{0}$
  in the rest frame of the $K^{0}$ and first $p$.

  \subsection{Scattering amplitude of the $K^{+} d \to K^{0} p p$ reaction} 
\label{sec:2-2}

\begin{figure}[htp]
  \centering
  \begin{tabular}{ccc}
    \PsfigII{0.16}{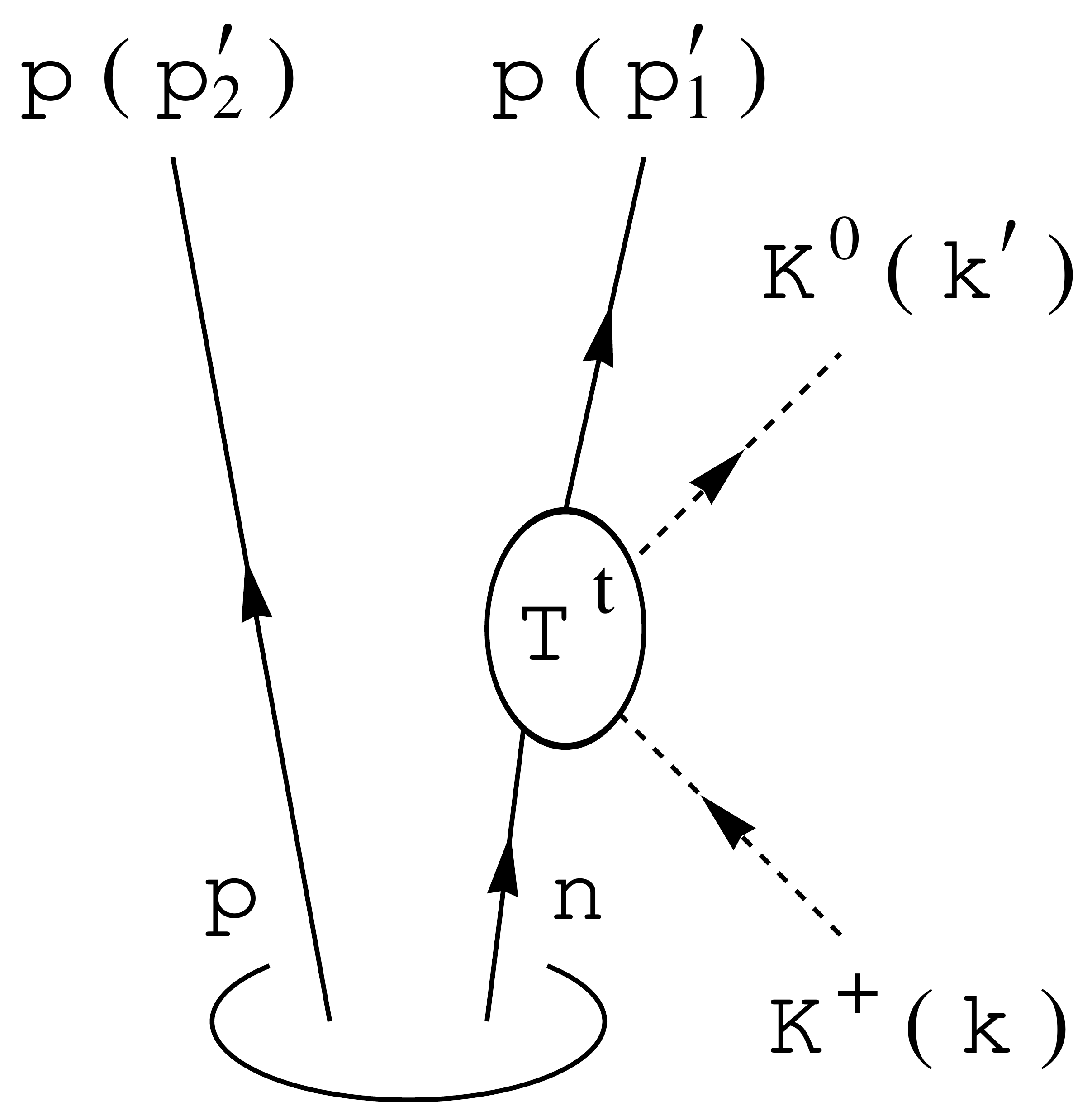} & \PsfigII{0.16}{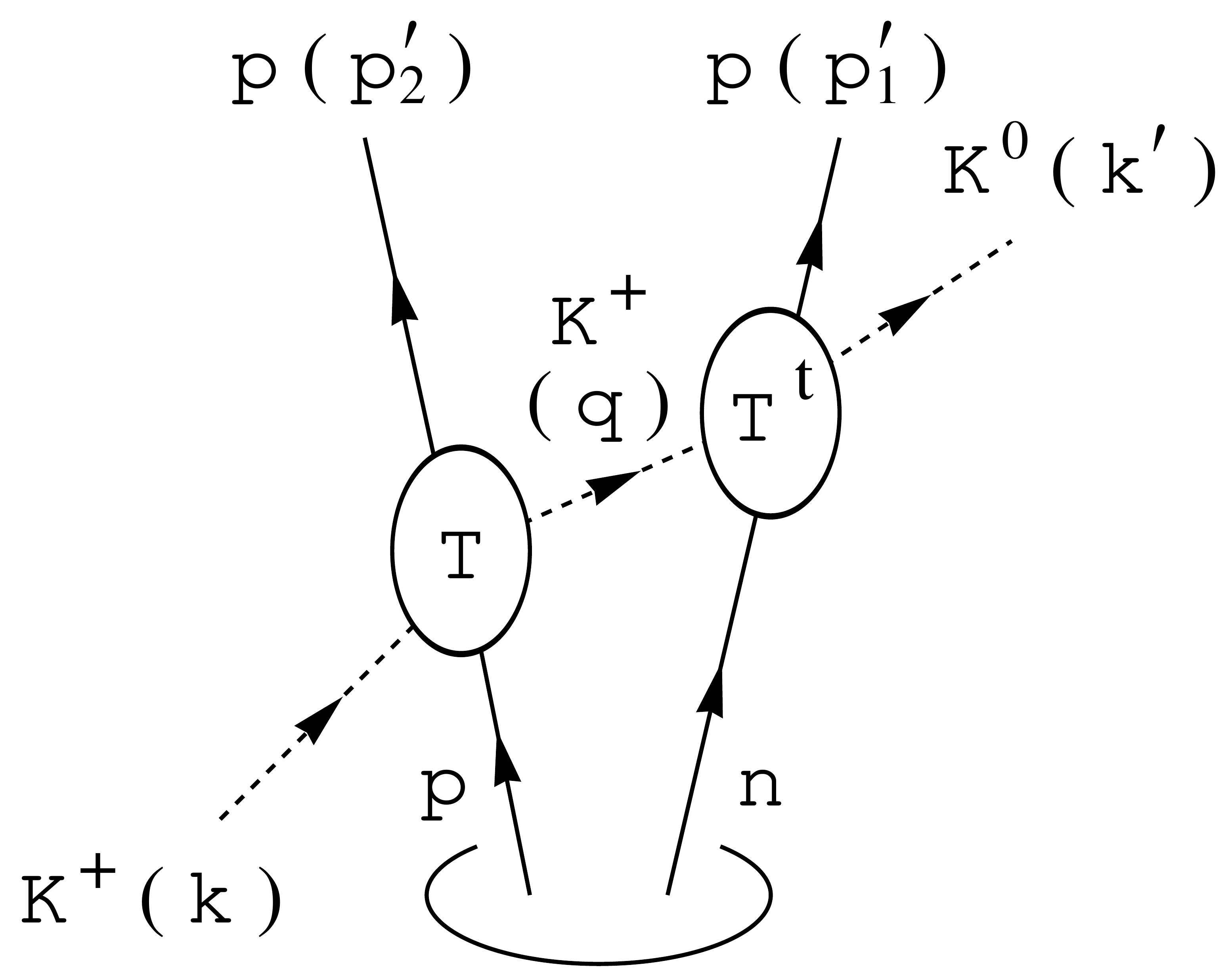} & \PsfigII{0.16}{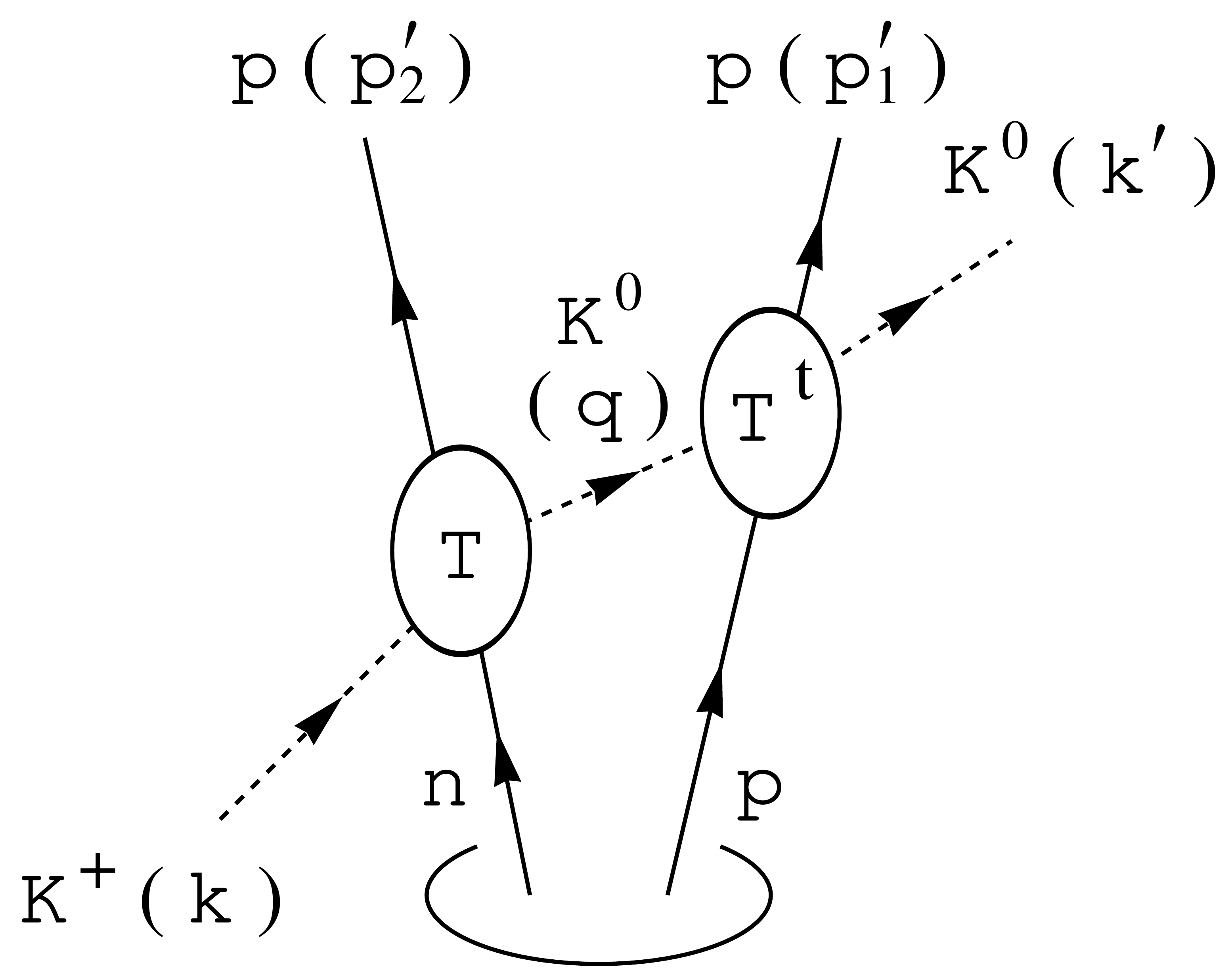}
    \\
    (a) $\mathcal{T}_{1}^{a}$ & (b) $\mathcal{T}_{2}^{a}$
    & (c) $\mathcal{T}_{3}^{a}$ 
  \end{tabular}
  \caption{Diagrams for the $K^{+} d \to K^{0} p p$ reaction.  Momenta
    of particles are shown in parentheses.}
  \label{fig:1}
\end{figure}

Next we construct the scattering amplitude of the $K^{+} d \to K^{0} p
p$ reaction. Since the deuteron has spin $1$, the scattering amplitude
can be denoted by $\mathcal{T}^{a}$ with $a=1,\,2$ and 3 to specify
the deuteron spin component. As depicted in Fig.~\ref{fig:1} together
with the momenta of particles, the reaction mechanism consists of the
three main diagrams: $\mathcal{T}_1^a$ stands for the impulse
scattering process~\cite{Sibirtsev:2004bg}, and $\mathcal{T}_2^a$ and
$\mathcal{T}_3^a$ represent the double-step scattering processes,
where the intermediate kaons $K^+$ and $K^0$ propagate respectively
between two nucleons\footnote{The double scattering contributions in
  the $K^{+} d$ reaction were taken into account in, e.g.,
  Ref.~\cite{Gibbs:2004ji} but in the case of lower kaon
  momenta.}. Thus, the $K^{+} d \to K^{0} p p$ scattering amplitude is
expressed as the sum of these three contributions:
\begin{align}
  \mathcal{T}^{a} =
  & \mathcal{T}_{1}^{a} ( k^{\mu} , \, k^{\prime \, \mu} , \,
  p_{1}^{\prime \, \mu} , \, p_{2}^{\prime \, \mu}  ) 
  +
  \mathcal{T}_{2}^{a} ( k^{\mu} , \, k^{\prime \, \mu} , \,
  p_{1}^{\prime \, \mu} , \, p_{2}^{\prime \, \mu}  ) 
  +
  \mathcal{T}_{3}^{a} ( k^{\mu} , \, k^{\prime \, \mu} , \,
  p_{1}^{\prime \, \mu} , \, p_{2}^{\prime \, \mu}  )
  \notag \\
  & - (\text{antisymmetric terms}) ,
\end{align}
where the antisymmetric terms are required owing to the identical
fermions, i.e., protons in the final state.  We now derive the
$K^{+} d$ scattering amplitude in the Lab frame, in which
the deuteron three-momentum satisfies $\bm{p}_{d} = \bm{0}$. 
In particular, we evaluate each $K N \to K N$ amplitude in the
target-baryon rest frame, as we will show below. 

The impulse scattering amplitude $\mathcal{T}_{1}^{a}$, depicted in
Fig.~\ref{fig:1}(a), is calculated
as~\cite{YamagataSekihara:2012yv}
\begin{equation}
  \mathcal{T}_{1}^{a} ( k^{\mu} , \, k^{\prime \, \mu} , \,
  p_{1}^{\prime \, \mu} , \, p_{2}^{\prime \, \mu}  )
  = \tilde{\varphi}
  ( | \bm{p}_{2}^{\prime}   | )  ( S^{\dagger} )^{a}
  T_{K^{+} n \to K^{0} p}^{\rm t} ( w_{1} ; \, \bm{k} , \, \bm{k}^{\prime} ) .
\end{equation}
Here, $T_{K^{+} n \to K^{0} p}$ stands for the $K^{+} n \to K^{0} p$
scattering amplitude in a $2 \times 2$ matrix form, which is
represented in the spin space of the nucleon, and the superscript
$\text{t}$ designates the transpose of a $2 \times 2$ matrix.  The
$K^{+} n \to K^{0} p$ amplitude depends on the CM energy $w_{1} =
\sqrt{( k^{\prime     \, \mu} + p_{1}^{\prime \, \mu} )^{2}}$ and
three-momenta of the initial and final kaons in the Lab frame,
$\bm{k}$ and $\bm{k}^{\prime}$, respectively.  The deuteron spin
component is denoted by $( S^{\dagger} )^{a} = - i \sigma ^{2} \sigma
^{a} / \sqrt{2}$ ($a = 1$, $2$, $3$) in a $2 \times 2$ matrix form
with the Pauli matrices $\sigma ^{a}$.  $\tilde{\varphi}$ is the
deuteron wave function in momentum space, for which we neglect the
$d$-wave component because it is negligibly small.  An analytic
parameterization of the $s$-wave component~\cite{Lacombe:1981eg}
facilitates the deutron wave function to be handled in an easy manner
\begin{equation}
  \tilde{\varphi} ( p ) = \sum _{j = 1}^{11} \frac{C_{j}}{p^{2} + m_{j}^{2}} ,
\end{equation}
with $C_{j}$ and $m_{j}$ determined in Ref.~\cite{Machleidt:2000ge}.
As we mentioned previously, each part of the $K^{+} d$ scattering
amplitude, i.e., the deuteron wave function and the $K^{+} n \to K^{0}
p$ amplitude, is evaluated in the Lab frame.  The expression of the
$K^{+} n \to K^{0} p$ amplitude in the target-baryon rest frame will
be given in Sec.~\ref{sec:2-3}.

The double-step scattering amplitudes, $\mathcal{T}_{2}^{a}$ and
$\mathcal{T}_{3}^{a}$, which are depicted respectively in
Fig.~\ref{fig:1}(b) and (c), are calculated
as~\cite{YamagataSekihara:2012yv} 
\begin{align}
  \mathcal{T}_{2}^{a} ( k^{\mu} , \, k^{\prime \, \mu} , \,
  p_{1}^{\prime \, \mu} , \, p_{2}^{\prime \, \mu}  )
  = & \int \frac{d^{3} q}{( 2 \pi )^{3}}
  \frac{\tilde{\varphi} ( | \bm{q} + \bm{p}_{2}^{\prime} - \bm{k}
    | )}{( q^{0} )^{2} - \bm{q}^{2} - m_{K^{+}}^{2} + i 0}
  F ( | \bm{q} | ) 
  \notag \\ & \times
  T_{K^{+} p \to K^{+} p} ( w_{2} ; \, \bm{k} , \, \bm{q} )
  ( S^{\dagger} )^{a}
  T_{K^{+} n \to K^{0} p}^{\rm t} ( w_{1} ; \, \bm{q} , \, \bm{k}^{\prime} ) ,
\end{align}
\begin{align}
  \mathcal{T}_{3}^{a} ( k^{\mu} , \, k^{\prime \, \mu} , \,
  p_{1}^{\prime \, \mu} , \, p_{2}^{\prime \, \mu}  )
  = & - \int \frac{d^{3} q}{( 2 \pi )^{3}}
  \frac{\tilde{\varphi} ( | \bm{q} + \bm{p}_{2}^{\prime} - \bm{k}
    | )}{( q^{0} )^{2} - \bm{q}^{2} - m_{K^{0}}^{2} + i 0}
  F ( | \bm{q} | ) 
  \notag \\ & \times
  T_{K^{+} n \to K^{0} p} ( w_{2} ; \, \bm{k} , \, \bm{q} )
  ( S^{\dagger} )^{a}
  T_{K^{0} p \to K^{0} p}^{\rm t} ( w_{1} ; \, \bm{q} , \, \bm{k}^{\prime} ) ,
\end{align}
where $F ( q )$ represents a form factor for which we take a Gaussian
form of $F ( q ) = \exp( - q^{2} / \Lambda ^{2} )$ with a cutoff
$\Lambda$.  The kaon energies in the propagators are fixed in the
truncated Faddeev approach~\cite{Jido:2012cy} as
\begin{equation}
  q^{0} = k^{0} + m_{d} - p_{2}^{\prime \, 0} 
  - \sqrt{| \bm{q} + \bm{p}_{2}^{\prime} - \bm{k} |^{2} + m_{N}^{2}} ,
\end{equation}
where the energy-momenta of the particles are fixed in the Lab frame
and $m_{N}$ is the isospin-averaged nucleon mass $m_N=(m_p+m_n)/2$.
The CM energy for the first collision is given by $w_{2} = \sqrt{(
  q^{\mu} + p_{2}^{\prime \, \mu} )^{2}}$, and hence it depends on the
Fermi motion of bound nucleons as well as on the initial-kaon
momentum.

As for the antisymmetric terms, we have to calculate the scattering
amplitudes where the momenta and spins of the two protons are
simultaneously exchanged, i.e., $( p_{1}^{\mu} , \, s_{1} )
\leftrightarrow (p_{2}^{\mu} , \, s_{2} )$ with $s_{1,2}$ being the
spins of protons. This antisymmetrization for the present scattering
amplitudes can be performed in the following manner:
\begin{equation}
  \left [ \mathcal{T}_{1,2,3}^{a} ( k^{\mu} , \, k^{\prime \, \mu} , \,
    p_{1}^{\prime \, \mu} , \, p_{2}^{\prime \, \mu}  ) \right ]
  _{\rm antisymmetric}
  = { \mathcal{T}_{1,2,3}^{a} }^{\text{t}} ( k^{\mu} , \, k^{\prime \, \mu} , \,
    p_{2}^{\prime \, \mu} , \, p_{1}^{\prime \, \mu}  ) .
\end{equation}

Finally, the squared scattering amplitude in Eq.~\eqref{eq:ds} is
obtained by the spin average and summation for the initial deuteron
and final protons, respectively, which results  
in the following expression~\cite{YamagataSekihara:2012yv}
\begin{equation}
  | \mathcal{T} | ^{2}
  = \frac{1}{3} \sum _{a = 1}^{3} \text{tr}
  \left [ \mathcal{T}^{a} \left ( \mathcal{T}^{a} \right )^{\dagger} \right ] .
\end{equation}

\subsection{Scattering amplitude of the $K N \to K N$ reaction}
\label{sec:2-3}

We turn to the $K N \to K N$ scattering amplitude $T_{K N \to K N}$,
which is expressed in terms of a $2 \times 2$ matrix. In the present
work, we first present the $K N \to K N$ amplitude in terms of the
partial waves in the CM frame of the $K N$ system, and then transform
it to that in the target-baryon rest frame, taking the method
in Ref.~\cite{Kamano:2016djv}.

The $K N$ amplitude is generally expressed in the $K N$ CM frame as:
\begin{align}
  & T_{K N \to K N}^{\rm cm}
  ( w ; \, \bm{p}_{\rm in}^{\ast} , \, \bm{p}_{\rm out}^{\ast} )
  \notag \\ & 
  = g_{K N \to K N}^{\rm cm}
  ( w , \, p_{\rm in}^{\ast} , \, p_{\rm out}^{\ast} , \, x^{\ast} )
  - i h_{K N \to K N}^{\rm cm}
  ( w , \, p_{\rm in}^{\ast} , \, p_{\rm out}^{\ast} , \, x^{\ast} )
  \frac{( \bm{p}_{\rm out}^{\ast} \times \bm{p}_{\rm in}^{\ast} ) \cdot \bm{\sigma}}
       {p_{\rm out}^{\ast} p_{\rm in}^{\ast}} ,
\end{align}
where $w$ denotes the CM energy and $\bm{p}_{\rm in}^{\ast}$
($\bm{p}_{\rm out}^{\ast}$) stands for the three-momentum for the
initial (final) kaon in the CM frame.  Then, we can define $p_{\rm
  out,in}^{\ast} \equiv | \bm{p}_{\rm out,in}^{\ast} |$ and $x^{\ast}
\equiv \bm{p}_{\rm   out}^{\ast} \cdot \bm{p}_{\rm in}^{\ast} / (
p_{\rm out}^{\ast} p_{\rm in}^{\ast} )$.  The Pauli matrices
$\bm{\sigma}$ act on the nucleon spinors, and $g_{K N \to K N}^{\rm
  cm}$ and $h_{K N \to K   N}^{\rm cm}$ are expressed in terms of the
partial waves as 
\begin{align}
  & g_{K N \to K N}^{\rm cm}
  ( w , \, p_{\rm in}^{\ast} , \, p_{\rm out}^{\ast} , \, x^{\ast} )
  \notag \\ &
  = \sum _{L = 0}^{\infty} \left [ ( L + 1 ) T_{K N \to K N , \, L +}^{\rm cm}
    ( w , \, p_{\rm in}^{\ast} , \, p_{\rm out}^{\ast} )
    + L T_{K N \to K N , \, L -}^{\rm cm}
    ( w , \, p_{\rm in}^{\ast} , \, p_{\rm out}^{\ast} )
    \right ]
  P_{L} ( x^{\ast} ) , \\
  & h_{K N \to K N}^{\rm cm}
  ( w , \, p_{\rm in}^{\ast} , \, p_{\rm out}^{\ast} , \, x^{\ast} )
  \notag \\
  & = \sum _{L = 1}^{\infty}
  \left [ T_{K N \to K N , \, L +}^{\rm cm}
    ( w  , \, p_{\rm in}^{\ast} , \, p_{\rm out}^{\ast} )
    - T_{K N \to K N, \, L -}^{\rm cm}
    ( w  , \, p_{\rm in}^{\ast} , \, p_{\rm out}^{\ast} ) \right ]
  P_{L}^{\prime} ( x^{\ast} ) ,
\end{align}
with the Legendre polynomials $P_{L} ( x )$, $P_{L}^{\prime} ( x )
\equiv d P_{L} / d x$, and orbital angular momentum $L$.

Next we transform the above-given amplitudes to that in the
target-baryon rest frame according to the
  formula~\cite{Kamano:2016djv}
  \begin{align}
    & T_{K N \to K N} ( w ; \, \bm{p}_{\rm in} , \, \bm{p}_{\rm out} )
    \notag \\
    & = \sqrt{\frac{\omega _{K} ( p_{\rm in}^{\ast} )
        E_{N} ( p_{\rm in}^{\ast} ) \omega _{K} ( p_{\rm out}^{\ast} )
        E_{N} ( p_{\rm out}^{\ast} )}
      {\omega _{K} ( p_{\rm in} ) m_{N} \omega _{K} ( p_{\rm out} )
        E_{N} ( | \bm{p}_{\rm in} - \bm{p}_{\rm out} | ) }}
    T_{K N \to K N}^{\rm cm}
    ( w ; \, \bm{p}_{\rm in}^{\ast} , \, \bm{p}_{\rm out}^{\ast} )
  \end{align}
  where parameters in the target-baryon rest frame are expressed
  without asterisks in contrast to those in the CM frame.
  The Lorentz-boost factor appears in the right-hand side and
  contains the kaon enerugy $\omega _{K} ( p ) \equiv \sqrt{m_{K}^{2}
    + p^{2}}$ with the isospin-averaged kaon mass $m_{K} = ( m_{K^{+}}
  + m_{K^{0}} ) / 2$ and nucleon enerugy $E_{N} ( p ) \equiv
  \sqrt{m_{N}^{2} + p^{2}}$.
 
We now construct the partial-wave amplitudes $T_{K N \to K N , \, L
  \pm}^{\rm cm}$, which should be in general off-shell amplitudes and
thus functions of three independent variables: $w$, $p_{\rm
  in}^{\ast}$, and $p_{\rm out}^{\ast}$.  In the present study, we
assume that the partial-wave amplitudes depend on the momenta
minimally required by the kinematics, i.e., the off-shell amplitudes
are proportional to $( p_{\rm out}^{\ast} p_{\rm in}^{\ast} )^{L}$.
Under this assumption, we have an advantage that the on-shell
amplitudes can simulate the off-shell amplitudes by introducing the
formula 
\begin{equation}
  T_{K N \to K N , \, L \pm}^{\rm cm}
  ( w , \, p_{\rm out}^{\ast} , \, p_{\rm in}^{\ast} )
  = T_{K N \to K N , \, L \pm}^{\text{on-shell}} ( w )
  \frac{( p_{\rm out}^{\ast} p_{\rm in}^{\ast} )^{L}}
       {[ p^{\text{on-shell}} ( w ) ]^{2 L}} ,
  \label{eq:min_on-shell}
\end{equation}
where $p^{\text{on-shell}}$ is the on-shell momentum for the
$K N$ system:
\begin{equation}
  p^{\text{on-shell}} ( w )
  = \frac{\lambda ^{1/2} ( w^{2} , \, m_{K}^{2} , \, m_{N}^{2} )}{2 w} .
\end{equation}
Since we need the $K N$ amplitudes in the energy range from its
threshold to $w \sim 2 \gev$, we utilize the on-shell $K N$ amplitude
developed in the SAID program~\cite{SAID}, in which they provide the
on-shell $K N$ amplitude in various partial waves.  We here take the
SAID partial-wave amplitudes up to the $D$ waves and calculate the
off-shell amplitudes by using Eq.~\eqref{eq:min_on-shell}. 

Finally we introduce the ``$\Theta ^{+}$'' contribution, which is just
added as an $s$-channel ``$\Theta ^{+}$'' exchange term to the $K N$
scattering amplitude in the present study.  Here we assume the 
``$\Theta ^{+}$'' as an isosinglet, and examine four different cases
of its spin/parity $J^{P} = 1/2^{\pm}$ and $3/2^{\pm}$.  The $K
N$``$\Theta $'' coupling is governed by an effective Lagrangian as
follows: 
\begin{equation}
  \mathcal{L} = g_{K N \Theta} \overline{\Theta} \Gamma
  ( K^{+} n - K^{0} p ) + \text{h.c.} ,
\end{equation}
in the spin $1/2$ case, where $\Gamma = 1$ ($i \gamma _{5}$) for the 
negative (positive) parity and $g_{K N \Theta}$ denotes the coupling
constant, and 
\begin{equation}
  \mathcal{L} = \frac{- i g_{K N \Theta}}{m_{K}} \overline{\Theta}^{\mu}
  \gamma _{5} \Gamma ( \partial _{\mu} K^{+} n - \partial _{\mu} K^{0} p )
  + \text{h.c.} ,
\end{equation}
in the spin $3/2$ case.  This provides us with the formula for the
``$\Theta ^{+}$'' decay width:
\begin{equation}
  \Gamma _{\Theta \to K^{0} p}
  = \Gamma _{\Theta \to K^{0} p}
  = \begin{cases}
    \displaystyle \frac{g_{K N \Theta}^{2} p_{K}^{\ast}
    \left ( E_{N} ( p_{K}^{\ast} ) \mp m_{N} \right )}
    {4 \pi M_{\Theta}}
    & \text{for } J^{P} = 1/2^{\pm} ,
    \\
    \displaystyle \frac{g_{K N \Theta}^{2} p_{K}^{\ast \, 3}
    \left ( E_{N} ( p_{K}^{\ast} ) \pm m_{N} \right )}
    {12 \pi m_{K}^{2} M_{\Theta}}
    & \text{for } J^{P} = 3/2^{\pm} ,
  \end{cases}
\end{equation}
where $M_{\Theta}$ stands for the ``$\Theta ^{+}$'' mass,
$p_{K}^{\ast}$ designates the CM momentum of the final-state $K N$
system. Thus, 
using the ``$\Theta ^{+}$'' mass $M_{\Theta}$ and full decay width
$\Gamma _{\Theta} = \Gamma _{\Theta \to K^{+} n} + \Gamma _{\Theta \to
  K^{0} p}$, we can fix the coupling constant $g_{K N \Theta}$.  In
the present study we use a presumable value of the mass $M_{\Theta} = 
1524 \mev$~\cite{Nakano:2008ee} and a predicted value of the decay
width $\Gamma _{\Theta} = 0.5 \mev$~\cite{Yang:2013tka}, which results
in $g_{K N \Theta} = 0.783$ for $J^{P} = 1/2^{+}$, $g_{K N \Theta} =
0.101$ for $1/2^{-}$, $g_{K N \Theta} = 0.352$ for $3/2^{+}$, and
$g_{K N \Theta} = 2.734$ for $3/2^{-}$.  Then, the $s$-channel
``$\Theta ^{+}$'' exchange term enters into the partial-wave $K^{+} n
\to K^{+} n$ amplitude as 
\begin{align}
  & T_{K^{+} n \to K^{+} n , \, 1 -}^{( \Theta )}
  ( w ; \, p_{\rm in}^{\ast} , \, p_{\rm out}^{\ast} )
  = \frac{p_{\rm out}^{\ast} p_{\rm in}^{\ast}}{4 m_{N}^{2}}
  \frac{g_{K N \Theta}^{2}}{w - M_{\Theta} + i \Gamma _{\Theta} / 2}
  & \text{for } J^{P} = 1/2^{+} ,
  \\
  & T_{K^{+} n \to K^{+} n , \, 0 +}^{( \Theta )}
  ( w ; \, p_{\rm in}^{\ast} , \, p_{\rm out}^{\ast} )
  =
  \frac{g_{K N \Theta}^{2}}{w - M_{\Theta} + i \Gamma _{\Theta} / 2}
  & \text{for } J^{P} = 1/2^{-} ,
  \\
  & T_{K^{+} n \to K^{+} n , \, 1 +}^{( \Theta )}
  ( w ; \, p_{\rm in}^{\ast} , \, p_{\rm out}^{\ast} )
  = \frac{p_{\rm out}^{\ast} p_{\rm in}^{\ast}}{3 m_{K}^{2}}
  \frac{g_{K N \Theta}^{2}}{w - M_{\Theta} + i \Gamma _{\Theta} / 2}
  & \text{for } J^{P} = 3/2^{+} ,
  \\
  & T_{K^{+} n \to K^{+} n , \, 2 -}^{( \Theta )}
  ( w ; \, p_{\rm in}^{\ast} , \, p_{\rm out}^{\ast} )
  = \frac{( p_{\rm out}^{\ast} p_{\rm in}^{\ast} )^{2}}{12 m_{K}^{2} m_{N}^{2}}
  \frac{g_{K N \Theta}^{2}}{w - M_{\Theta} + i \Gamma _{\Theta} / 2}
  & \text{for } J^{P} = 3/2^{-} .
\end{align}
The ``$\Theta ^{+}$'' contributions to the $K^{+} n \to K^{0} p$ and
$K^{0} p \to K^{0} p$ amplitudes are evaluated with the isospin
relation $T_{K^{+} n \to K^{0} p}^{( \Theta )} = - T_{K^{0} p \to
  K^{0} p}^{( \Theta )} = - T_{K^{+} n \to K^{+} n}^{( \Theta )}$.

\section{Numerical results and discussion}
\label{sec:3}
We are now in a position to present the numerical results and discuss
their physical implications, in particular how the signal of the
``$\Theta ^{+}$'' pentaquark, if it exists, emerges in the $K^{0} p$
spectrum.

\subsection{Two mechanisms to reach the ``$\Theta ^{+}$'' energy}

Before we discuss the details of the $K^{0} p$ spectrum, we first
examine which conditions are better to reach the ``$\Theta ^{+}$''
energy region and to search for its peak. To this end, we calculate
the $K^{0} p$ invariant mass of the $K^{+} d \to K^{0} p p$ reaction,
assuming that nucleon Fermi motion is zero and that the reaction takes
place only in the impulse scattering process.  In this case, one
proton is produced from a zero-momentum neutron in the initial state
of the $K^{+} n \to K^{0} p$ reaction, while the other proton comes
out just as a spectator.  Then, we can calculate the $K^{0} p$
invariant mass in two ways: combining $K^{0}$ with the produced proton
in the impulse scattering process, and doing $K^{0}$ with the
spectator proton.  Namely, in terms of the formulation in
Sec.~\ref{sec:2-1}, the former (latter) case means that the produced
proton is the ``first'' (``second'') proton while the spectator is the
``second'' (``first'') proton.

\begin{figure}[htp]
  \centering
  \Psfig{8.6cm}{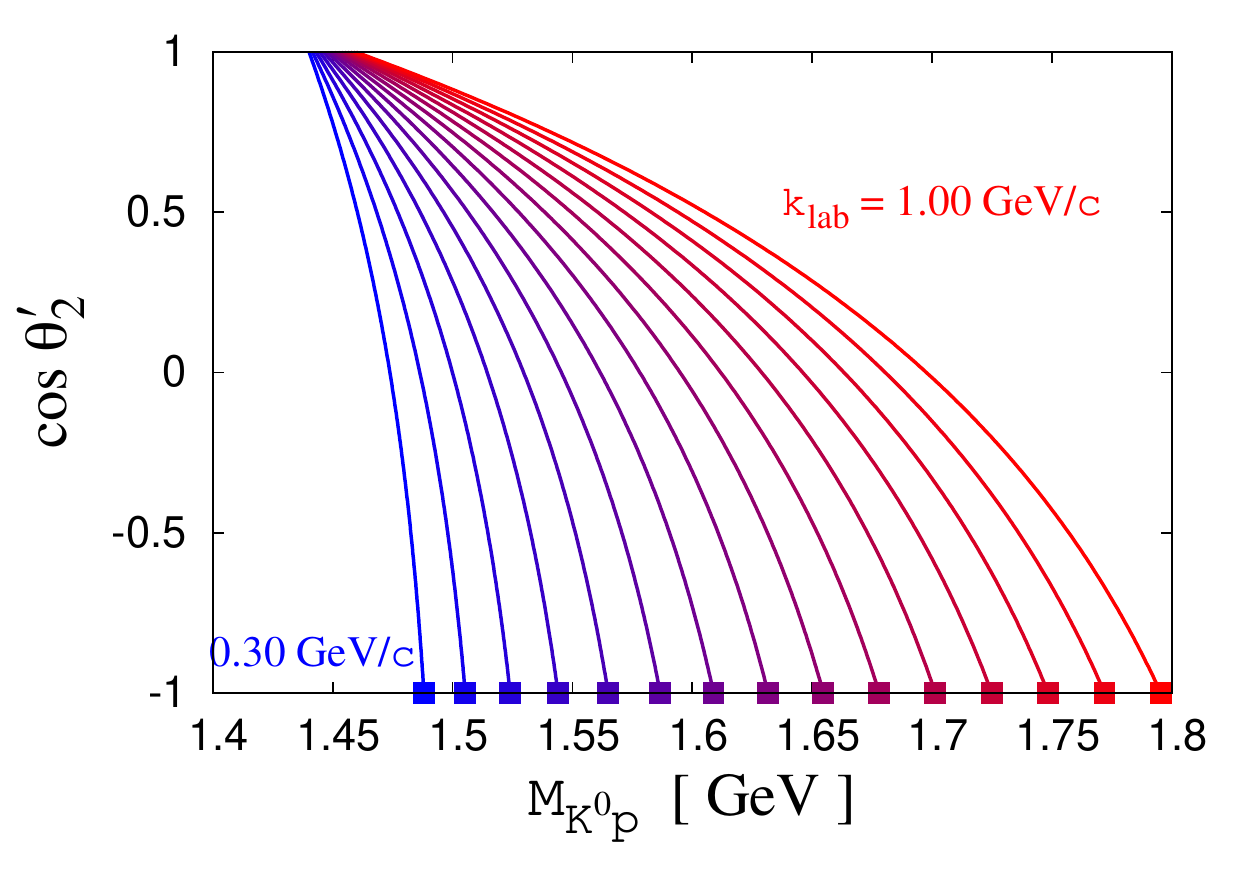}
  \caption{Possible $K^{0} p$ invariant mass of the $K^{+} d \to K^{0}
    p p$ reaction with the impulse scattering process as a function of
    the scattering angle for the ``second'' proton in the global CM
    frame $\theta _{2}^{\prime}$.  The initial kaon momentum in the
    Lab frame $k_{\rm lab}$ is taken from $0.30 \gev /c$ to $1.00 \gev
    /c$ in intervals of $0.05 \gev /c$.  We assume zero Fermi motion.
    The squared boxes represent the invariant mass of $K^{0}$ and
    produced proton of the impulse process, while the lines represent
    that of $K^{0}$ and spectator proton.}
  \label{fig:2}
\end{figure}

The $K^{0} p$ invariant mass in these two ways is plotted in
Fig.~\ref{fig:2} in terms of the squared boxes (former) and curves
(latter), respectively.  We note that, in general, the possible $K^{0}
p$ invariant mass discussed here is slightly smeared compared with
those in Fig.~\ref{fig:2} due to the Fermi motion of the nucleons
inside a deuteron.  As one can see, on the one hand, the squared boxes in
Fig.~\ref{fig:2} reaches around the ``$\Theta ^{+}$'' energy region
$\sim 1.52 \gev$ with the initial-kaon momentum $k_{\rm lab} \approx
0.40 \gev /c$.  This means that one can investigate the ``$\Theta
^{+}$'' energy region directly in the impulse scattering process with
the initial kaon momentum $k_{\rm lab} \approx 0.40 \gev /c$ and with
the backward ``second'' proton.  On the other hand, the curves in
Fig.~\ref{fig:2} suggest that even with higher kaon momenta one can
reach the ``$\Theta ^{+}$'' energy region by observing the forward
``second'' proton.  In this case, although the impulse scattering
process cannot produce the ``$\Theta ^{+}$'', the double-step
scattering one can do it, where the intermediate kaon
may lead to the formation of the ``$\Theta ^{+}$'', being combined
with the ``first'' proton.

Therefore, the present study consists of two folds.  Firstly, we check
whether a possible ``$\Theta ^{+}$'' signal will appear in the impulse
scattering process with lower kaon momenta $k_{\rm lab} \approx 0.40
\gev /c$. In fact, Ref.~\cite{Sibirtsev:2004bg} already carried it
out. Thus, in the first part of the present work, we extend
Ref.~\cite{Sibirtsev:2004bg} and perform a more detailed analysis of
the study. Secondly, we investigate whether the double-step scattering
contribution with higher kaon momenta $k_{\rm lab} \sim 1 \gev /c$  
can generate the ``$\Theta ^{+}$'' in the $K^{+} d \to K^{0} p p$
reaction.

\subsection{Lower kaon momentum}

Let us first consider the case of lower kaon momenta.  In this case,
we expect that impulse scattering process dominates the $K^{+} d
\to K^{0} p p$ reaction. Indeed, we will see 
that this is
the case at lower kaon momenta, since the double-step processes give
only a few percent contribution to the differential cross section.
In the following discussions for lower momentum
we examine the impulse scattering process only unless explicitly
  mentioned.  This allows one to check how the results of the cross
section for the $K^{+} d \to K^{0} p p$ reaction are affected when the
``$\Theta ^{+}$'' is taken into account.

\begin{figure}[htp]
  \centering
  \includegraphics[width=7.6cm]{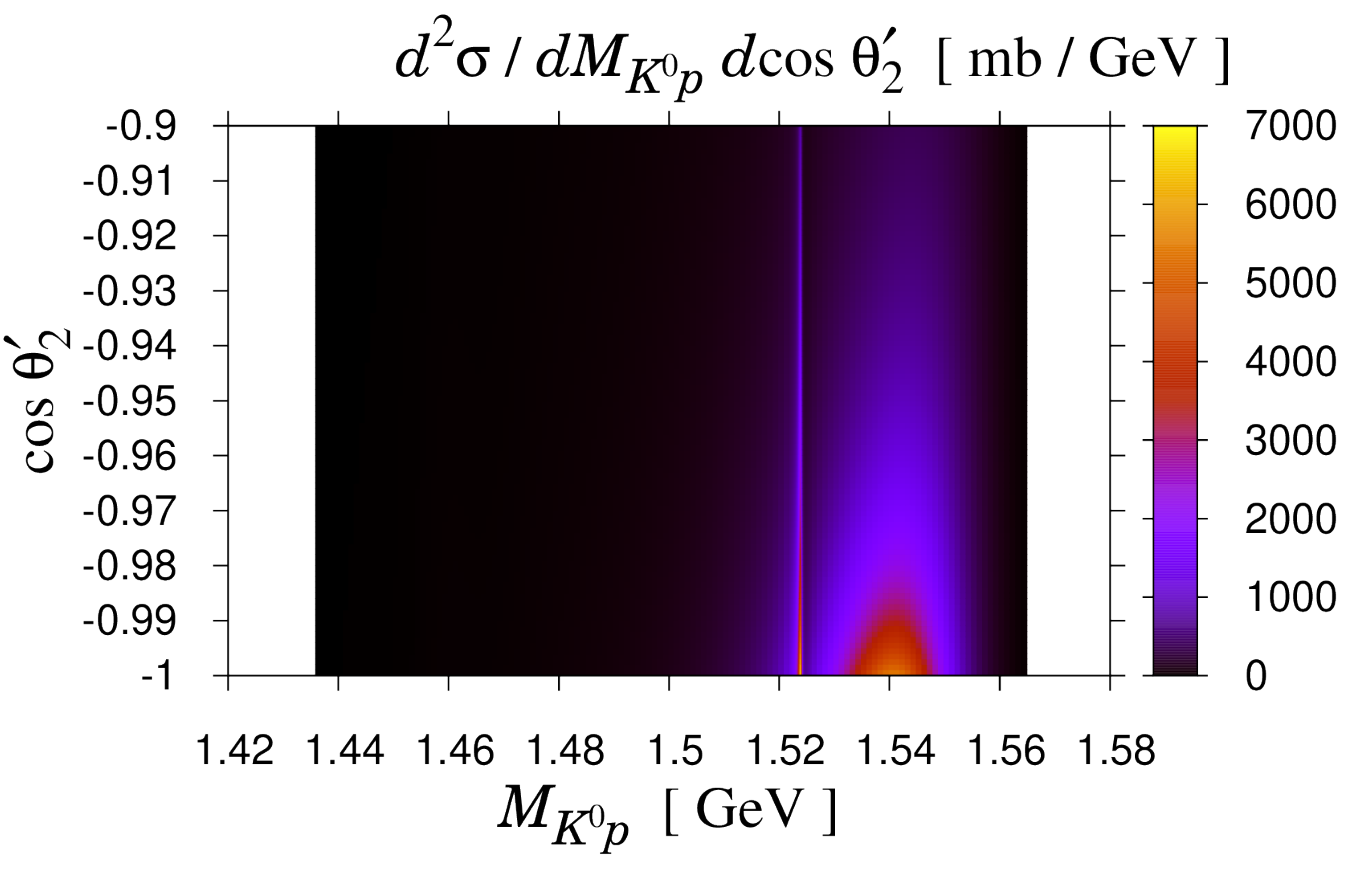}
~\includegraphics[width=7.0cm]{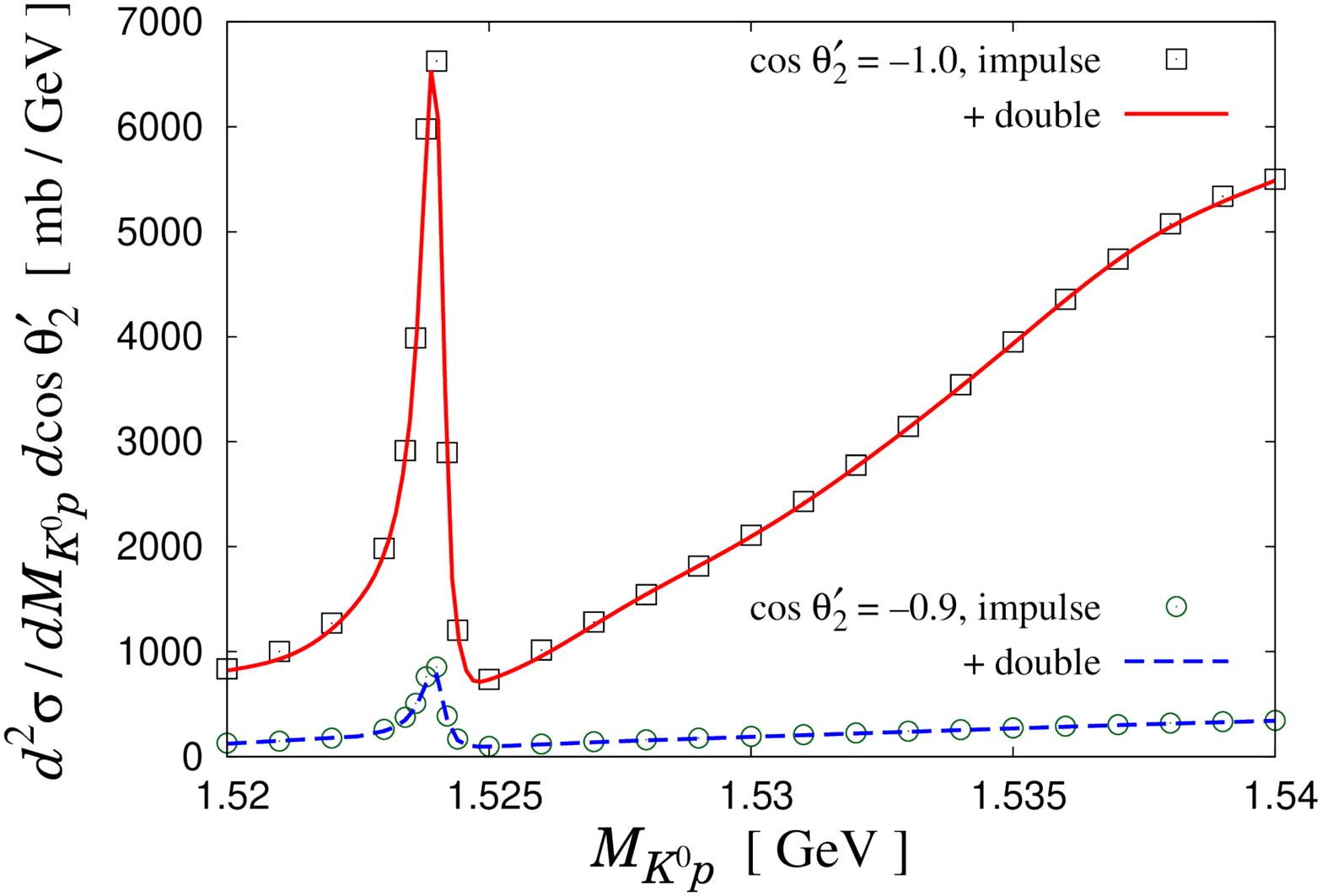}
  \caption{Results for the differential cross section of the $K^{+} d
    \to K^{0} p p$ reaction in the impulse scattering process
    (left) and the comparison of the differential cross section
      with and without the double-step processes at $\cos \theta
      _{2}^{\prime} = -1.0$ and $-0.9$ (right).  The initial kaon
    momentum is fixed as $k_{\rm lab} = 0.45 \gev /c$.  The
    ``$\Theta^{+}$'' with $J^{P} = 1/2^{+}$ is taken into account.}
  \label{fig:3}
\end{figure}

In order to see how the ``$\Theta ^{+}$'' influences the cross
section, we draw in Fig.~\ref{fig:3} the results for the differential
cross section $d^{2} \sigma / d M_{K^{0} p} \cos \theta _{2}^{\prime}$
with the initial kaon momentum $k_{\rm lab} = 0.45 \gev /c$.  The
``$\Theta ^{+}$'' with $J^{P} = 1/2^{+}$ is taken into account.  Here
we draw only the region $\cos \theta _{2}^{\prime} \le - 0.9$ because
there is no significant structure in the region $\cos \theta
_{2}^{\prime} > 0.9$.  We find two structures in the contour plot of
Fig.~\ref{fig:3}: a sharp peak at $M_{K^{0} p} = 1.524 \gev$ as a
``$\Theta ^{+}$'' signal and a broad bump at $M_{K^{0} p} = 1.54 \gev$
and $\cos \theta _{2}^{\prime} = -1$ corresponding to the squared
boxes in Fig.~\ref{fig:2}, arising from the kinematical effects.  The
Fermi motion of the bound neutron due to the deuteron wave function,
on the one hand, makes the peak at $( M_{K^{0} p} , \, \cos \theta
_{2}^{\prime} ) = ( 1.54 \gev , \, -1)$ broad.  On the other hand,
thanks to the same Fermi motion of the bound neutron, we can reach the
``$\Theta ^{+}$'' energy in the impulse scattering process even when
the initial kaon momentum does not exactly match the kaon momentum
that generates the two-body CM energy $1.524 \gev$ with a free nucleon
at rest, i.e., the kaon momentum $k_{\rm lab} \approx 0.40 \gev /c$.
Therefore, we can observe the ``$\Theta ^{+}$'' signal with $k_{\rm
  lab} = 0.45 \gev /c$ as in the left panel of Fig.~\ref{fig:3}.

To confirm that the impulse scattering process dominates the
  $K^{+} d \to K^{0} p p$ reaction in the lower kaon momentum case, we
  compare in the right panel of Fig.~\ref{fig:3} the differential
  cross section $d^{2} \sigma / d M_{K^{0} p} \cos \theta
  _{2}^{\prime}$ only with the impulse scattering (points) and that
  with the double-step processes (lines).  The initial kaon momentum
  is $k_{\rm lab} = 0.45 \gev /c$, the spin/parity of the ``$\Theta
  ^{+}$'' is $J^{P} = 1/2^{+}$, and the scattering angles $\cos \theta
  _{2}^{\prime} = -1.0$ and $-0.9$ are considered.  From the right
  panel of Fig.~\ref{fig:3} we can see that the contributions from the
  double-step processes are indeed negligible in the lower kaon
  momentum case.  Indeed, the double-step processes give only a few
  percent contribution to the differential cross section.  The same
  behavior is observed at other angles $\cos \theta _{2}^{\prime}$ and
  other (but lower) kaon momentum.  Therefore, we can safely
  concentrate on the impulse scattering process in this subsection.

\begin{figure}[htp]
  \centering
  \Psfig{10cm}{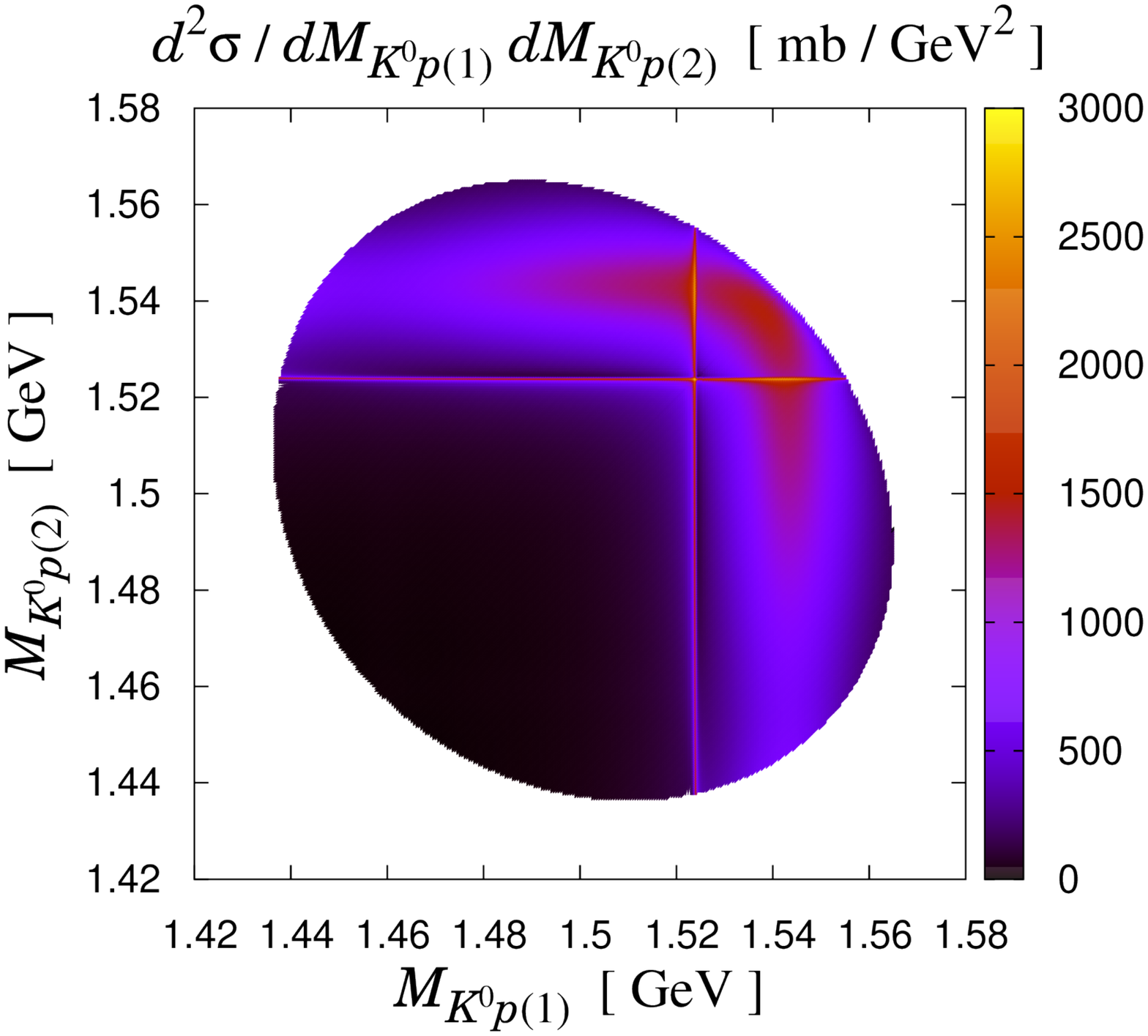}
  \caption{Dalits plot of the $K^{+} d \to K^{0} p p$ reaction in the
    impulse scattering process.  The initial kaon momentum is fixed as
    $k_{\rm lab} = 0.45 \gev /c$.  The ``$\Theta^{+}$'' with $J^{P} =
    1/2^{+}$ is taken into account.}
  \label{fig:4}
\end{figure}

We discuss the same reaction with the Dalitz plot $d^{2}
\sigma / d M_{K^{0} p (1)} d M_{K^{0} p (2)}$.  In
Fig.~\ref{fig:4} we show the Dalitz plot with $k_{\rm lab} = 0.45
\gev / c$ and the ``$\Theta ^{+}$'' with spin and parity $J^{P} =
1/2 ^{+}$. Broad structures around $M_{K^{0} p (1)} \sim 1.54$ in
the vertical direction and $M_{K^{0} p (2)} \sim 1.54$ in the horizontal
direction originate from the kinematical effects, which correspond to
the squared boxes in Fig.~\ref{fig:2}.  Besides, sharp structures at
$M_{K^{0} p (1)} = 1.524$ in the vertical direction and $M_{K^{0}
  p(2)} = 1.524$ in the horizontal direction indicate the
``$\Theta^{+}$'' signal.

\begin{figure}[htp]
  \centering
  \Psfig{8.6cm}{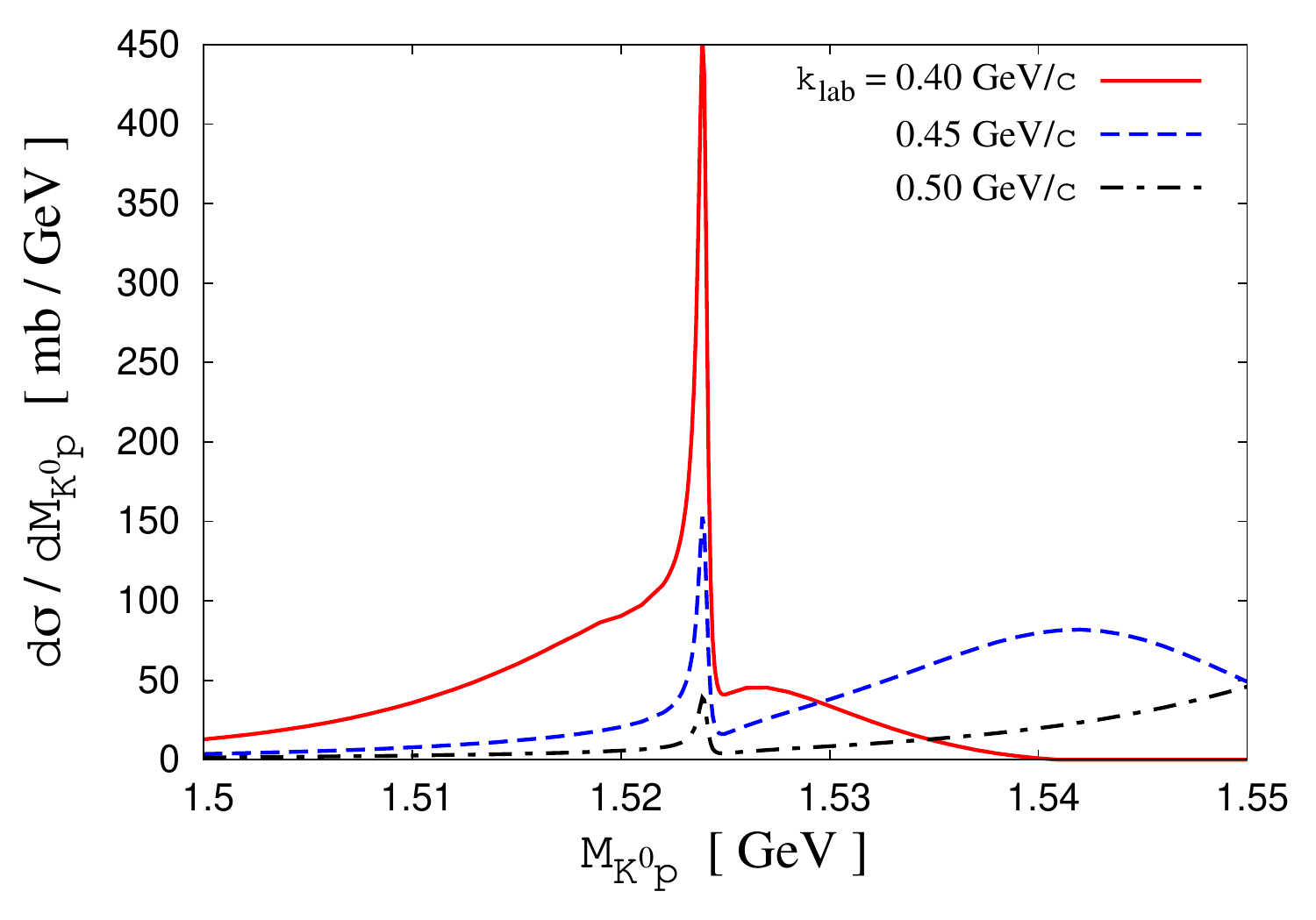}
  \caption{$K^{0} p$ invariant mass spectrum of the $K^{+} d \to K^{0}
    p p$ reaction with the initial kaon momentum $k_{\rm lab} = 0.40
    \gev /c$, $0.45 \gev /c$, and $0.50 \gev / c$ in the impulse
    scattering process.  Integral range of the scattering angle is $-1
    < \cos \theta _{2}^{\prime} < -0.8$.  The ``$\Theta ^{+}$'' with
    $J^{P} = 1/2^{+}$ is taken into account.}
  \label{fig:5}
\end{figure}

In Fig.~\ref{fig:5} we plot the $K^{0} p$ invariant mass
spectra of the $K^{+} d \to K^{0} p p$ reaction with three
initial-kaon momenta $k_{\rm lab} = 0.40 \gev /c$, $0.45 \gev / c$,
and $0.50 \gev /c$.  Here we take into account the ``$\Theta ^{+}$''
contribution with $J^{P} = 1/2^{+}$, and we integrate with respect to
the scattering angle in the range $-1 < \cos \theta _{2}^{\prime} <
-0.8$.  As one can see from Fig~\ref{fig:5}, on the one hand, the
broad-peak structure, which corresponds to impulse scattering of the
initial kaon and almost on-shell bound neutron, moves upward as
$k_{\rm lab}$ increases, as expected from the squared boxes in
Fig.~\ref{fig:2}.  On the other hand, the ``$\Theta ^{+}$'' signal
stays at $1.524 \gev$ with different values of $k_{\rm lab}$.  Among
three values of the kaon momenta, $k_{\rm lab} = 0.40 \gev / c$
yields the highest peak at $M_{K^{0} p} = 1.524 \gev$ for the
``$\Theta ^{+}$'' signal on top of the broad peak.  This is a
consequence of the momentum matching in Fig.~\ref{fig:2}.

\begin{figure}[htp]
  \centering
  \Psfig{8.6cm}{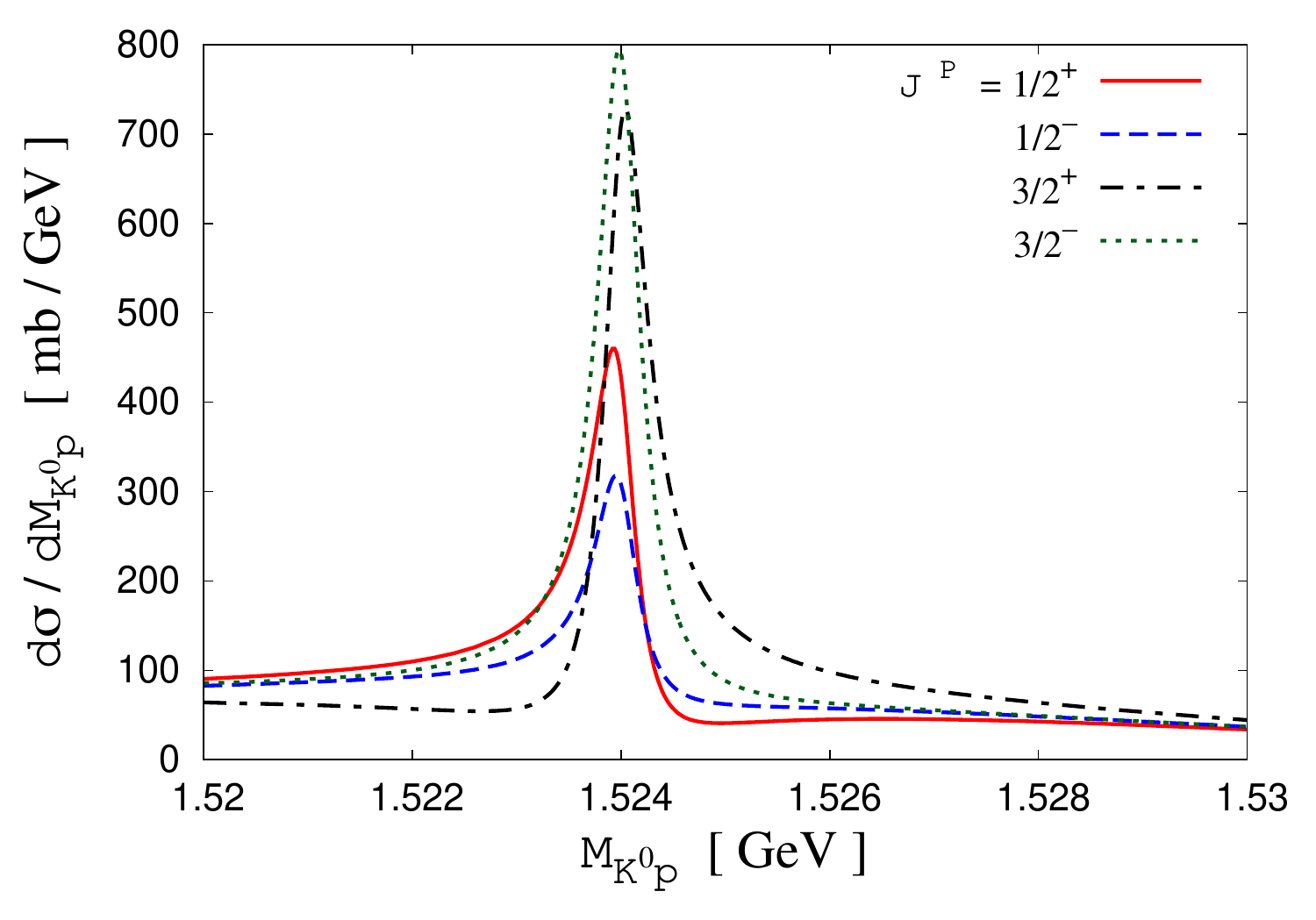}
  \caption{$K^{0} p$ invariant mass spectrum of the $K^{+} d \to K^{0}
    p p$ reaction in the impulse scattering scattering with the
    ``$\Theta ^{+}$'' spin/parity $J^{P} = 1/2^{\pm}$ and $3/2^{\pm}$.
    The initial kaon momentum is fixed as $k_{\rm lab} = 0.40 \gev
    /c$.  Integral range of the scattering angle is $-1 < \cos \theta
    _{2}^{\prime} < -0.8$.}
  \label{fig:6}
\end{figure}

We then examine other spin/parity combinations of the ``$\Theta ^{+}$''
pentaquark: $J^{P} = 1/2^{-}$, $3/2^{+}$, and $3/2^{-}$.  In
Fig.~\ref{fig:6} we show the $K^{0} p$ invariant mass spectrum
of the $K^{+} d \to K^{0} p p$ reaction with the initial kaon momentum
$k_{\rm lab} = 0.40 \gev / c$ and with the ``$\Theta ^{+}$'' of $J^{P}
= 1/2^{\pm}$ and $3/2^{\pm}$.  The integral range of the scattering
angle is $-1 < \cos \theta _{2}^{\prime} < -0.8$.  Figure~\ref{fig:6}
indicates that the peak height for the ``$\Theta ^{+}$'' signal in different
quantum numbers is similar to each other.  These peaks generate the
``$\Theta ^{+}$'' production cross section $\sim$ several hundred $\mu
\text{b}$ to $1 \millib$ with $k_{\rm lab} \approx 0.40 \gev / c$.

\begin{figure}[htp]
  \centering
  \Psfig{8.6cm}{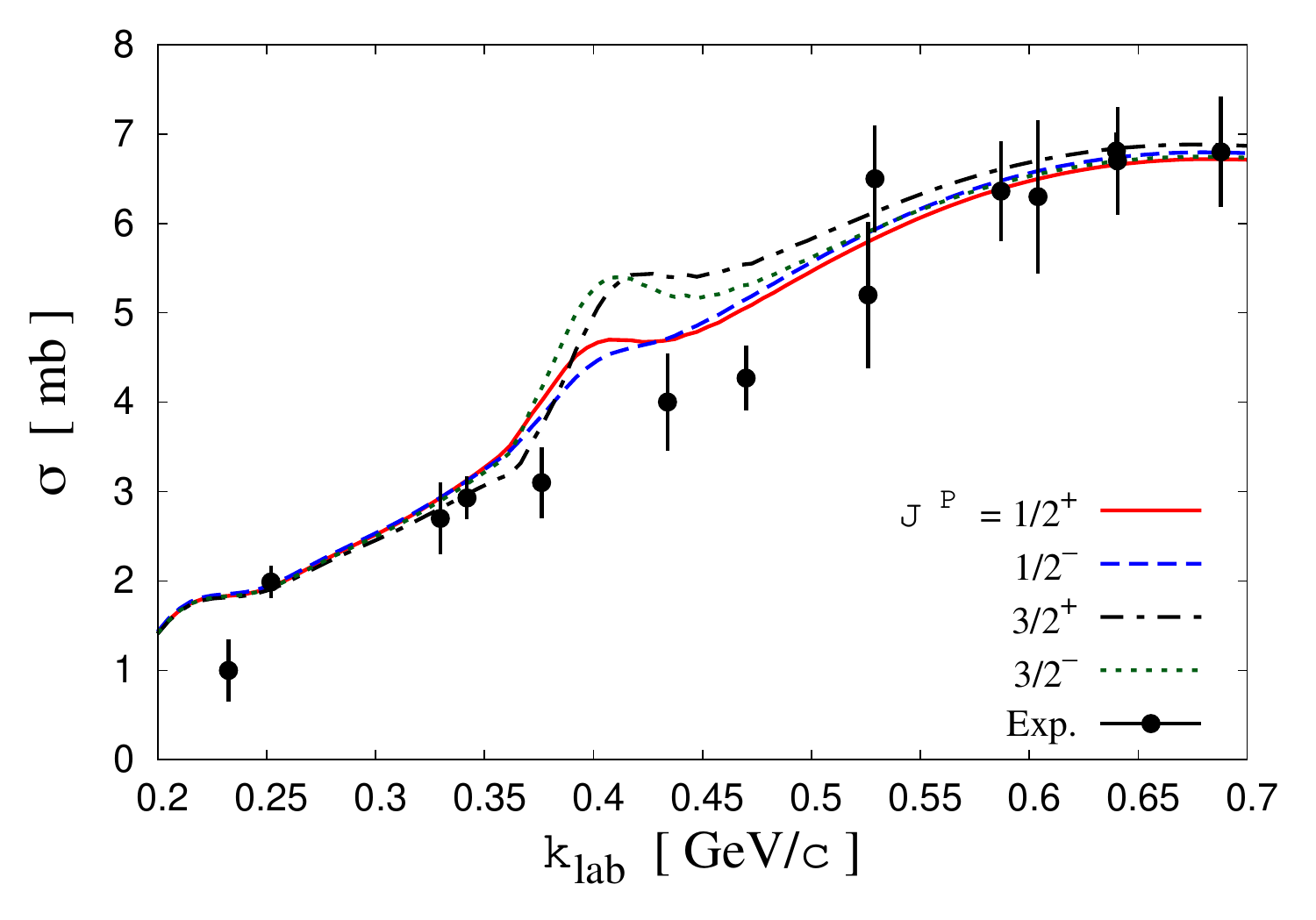}
  \caption{Total cross section of the $K^{+} d \to K^{0} p p$ reaction
    as a function of the initial kaon momentum in the Lab frame
    $k_{\rm lab}$.  We take into account only the impulse scattering.
    The experimental data are taken from Refs.~\cite{Slater:1961zz,
      Giacomelli:1972vb, Damerell:1975kw, Glasser:1977xs}.}
  \label{fig:7}
\end{figure}

Finally we calculate the total cross section of the $K^{+} d \to K^{0}
p p$ reaction with the ``$\Theta ^{+}$'' contribution of spin/parity
$J^{P} = 1/2^{\pm}$ and $3/2^{\pm}$. The result is shown in
Fig.~\ref{fig:7} together with the old experimental data on $K^+d\to
K^0pp$ scattering~\cite{Slater:1961zz, Giacomelli:1972vb,
  Damerell:1975kw, Glasser:1977xs}. 
Note that similar results were already obtained in
Ref.~\cite{Sibirtsev:2004bg} in which however the width of the
``$\Theta^+$'' was taken to be $1$--$20 \mev$.  As shown in
Fig.~\ref{fig:7}, even if the decay width of the ``$\Theta ^{+}$'' is
as small as $\Gamma _{\Theta} = 0.5 \mev$, which is approximately two
to forty times smaller than those in Ref.~\cite{Sibirtsev:2004bg}, one
can observe a bump structure around the initial kaon momentum in the
Lab frame $k_{\rm lab} = 0.4 \gev / c$.  The height of the bump gives
indeed a few hundred $\mu \text{b}$ to $1 \millib$. While the old
experiments have a lack of the data in the vicinity of
$k_{\mathrm{lab}} \approx 0.4\,\gev/c$, new experiments at the J-PARC,
if it is performed exclusively near this value of the initial kaon
momentum in the near future, can judge the existence of the
``$\Theta^+$'', because the size of the bump structure (a few hundred
$\mu \text{b}$ to $1 \millib$) is still strong enough to be seen.
In order to check a bump structure in the total cross section in
Fig.~\ref{fig:7}, a required resolution of the initial kaon momentum
is about several ten MeV$/ c$, while one can observe the ``$\Theta
^{+}$'' peak, if it exists, in the $K^{0} p$ invariant mass spectrum
 as in Fig.~\ref{fig:6} with the resolution of the $K^{0} p$
 invariant mass $\sim 1 \mev$.

\subsection{Higher kaon momentum}
We now focus on the case of higher kaon momenta.  
To reach the ``$\Theta ^{+}$'' energy region with higher kaon
momenta, we need to consider the double-step scattering process 
where the initial $K^+$ produces a proton from the deuteron in the
first collision, losing some part of its momentum. Then, it interacts
with the other nucleon in the second collision.  In this process, the first
collision corresponds to the $K^{+} p \to K^{+} p$ or $K^{+} n \to
K^{0} p$ reaction of the forward proton emission.  In this sense,
the initial kaon momentum should be chosen such that the forward
proton emission efficiently takes place in the 
first collision.  In other words, we require a specific initial kaon
momentum in such a way that the $K^{+} p \to K^{+} p$ and $K^{+} n \to
K^{0} p$ cross sections with the forward proton emission should be
large. In fact, this strategy was employed to search for a $\bar{K} N
N$ quasi-bound state in the $K^{-} {}^{3}\text{He} \to \Lambda p n$
reaction in the J-PARC E15 experiment~\cite{Sada:2016nkb,
  Ajimura:2018iyx}.  In the J-PARC E15 experiment, to generate a
$\bar{K} N N$ quasi-bound state, they planned to prepare a slow
antikaon and two of the three bound nucleons in a ${}^{3} \text{He}$
by using the $K^{-} n \to K^{-} n$ or $K^{-} p \to \bar{K}^{0} n$ reaction
with the fast forward neutron emission as the first collision, which
eventually leads to the $\bar{K} N N$  quasi-bound state (see also a
theoretical calculation of the $K^{-}{}^{3}\text{He} \to \Lambda p n$
reaction in Ref.~\cite{Sekihara:2016vyd}).  To prepare a slow  
antikaon and fast forward neutron emission as much as possible, 
the initial $K^{-}$ momentum $k_{\rm lab} = 1.0 \gev /c$ was selected
in Refs.~\cite{Sada:2016nkb, Ajimura:2018iyx}.

\begin{figure}[htp]
  \centering
  \Psfig{7.6cm}{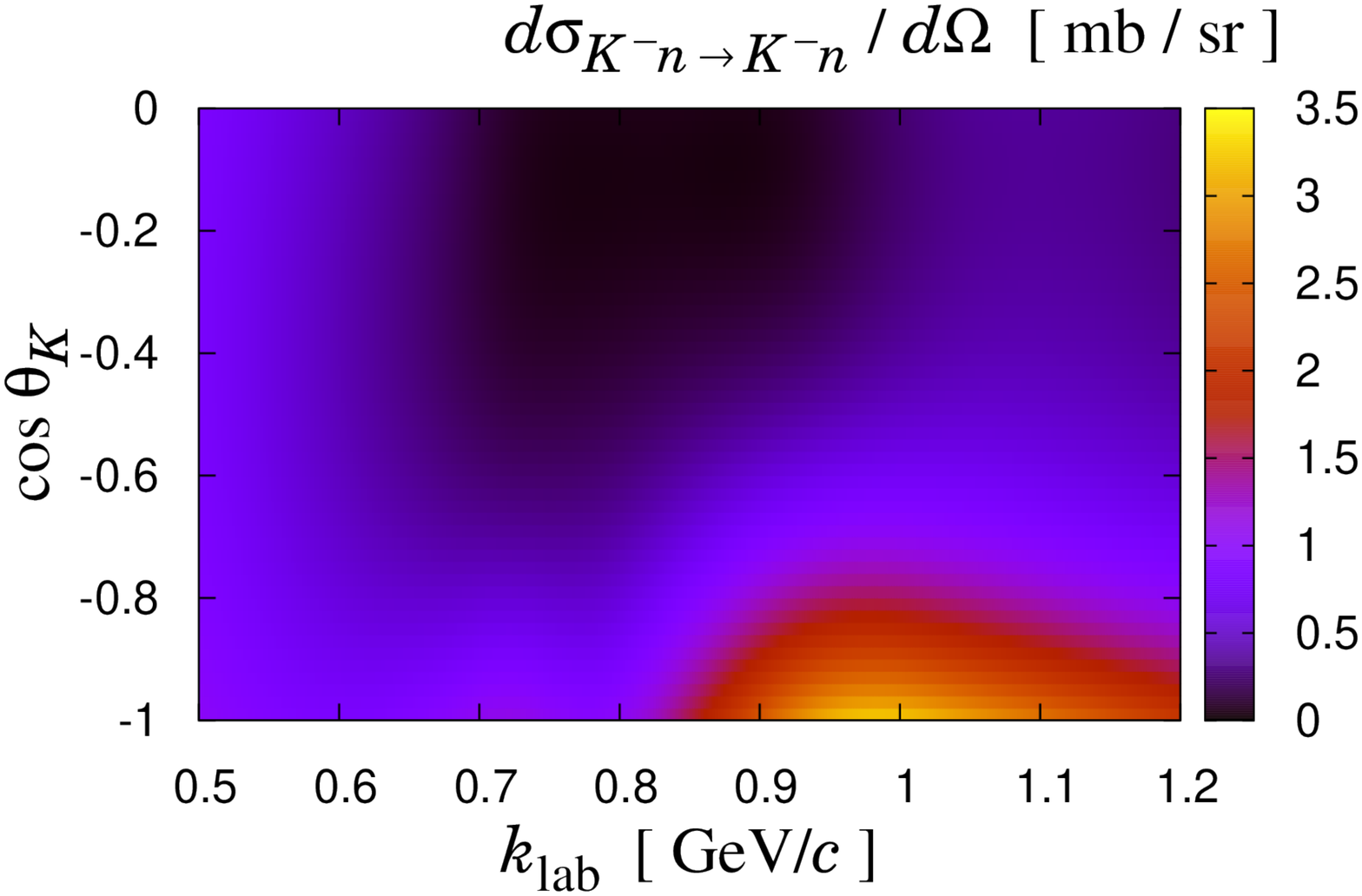}
  \Psfig{7.6cm}{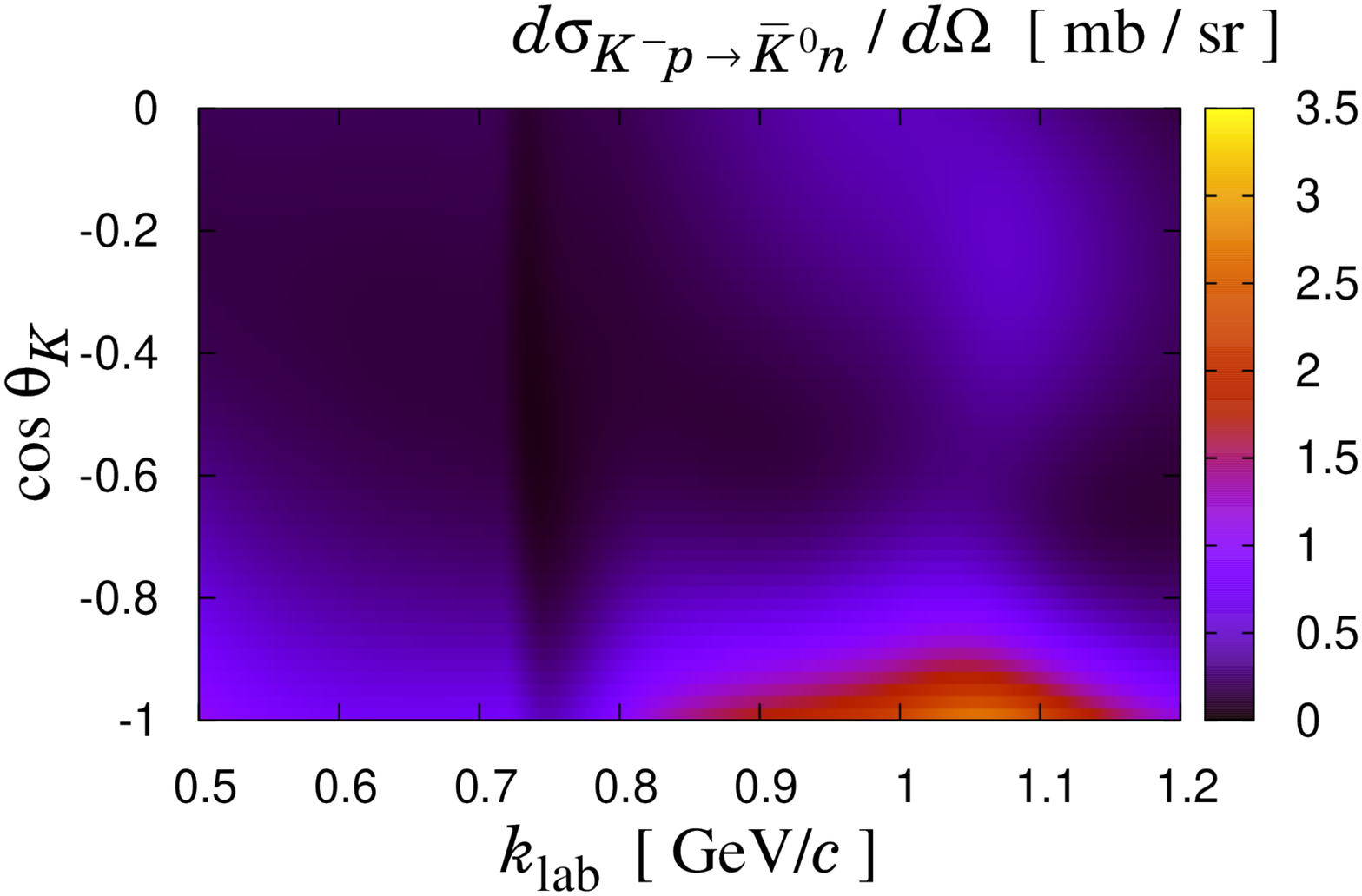}
  \Psfig{7.6cm}{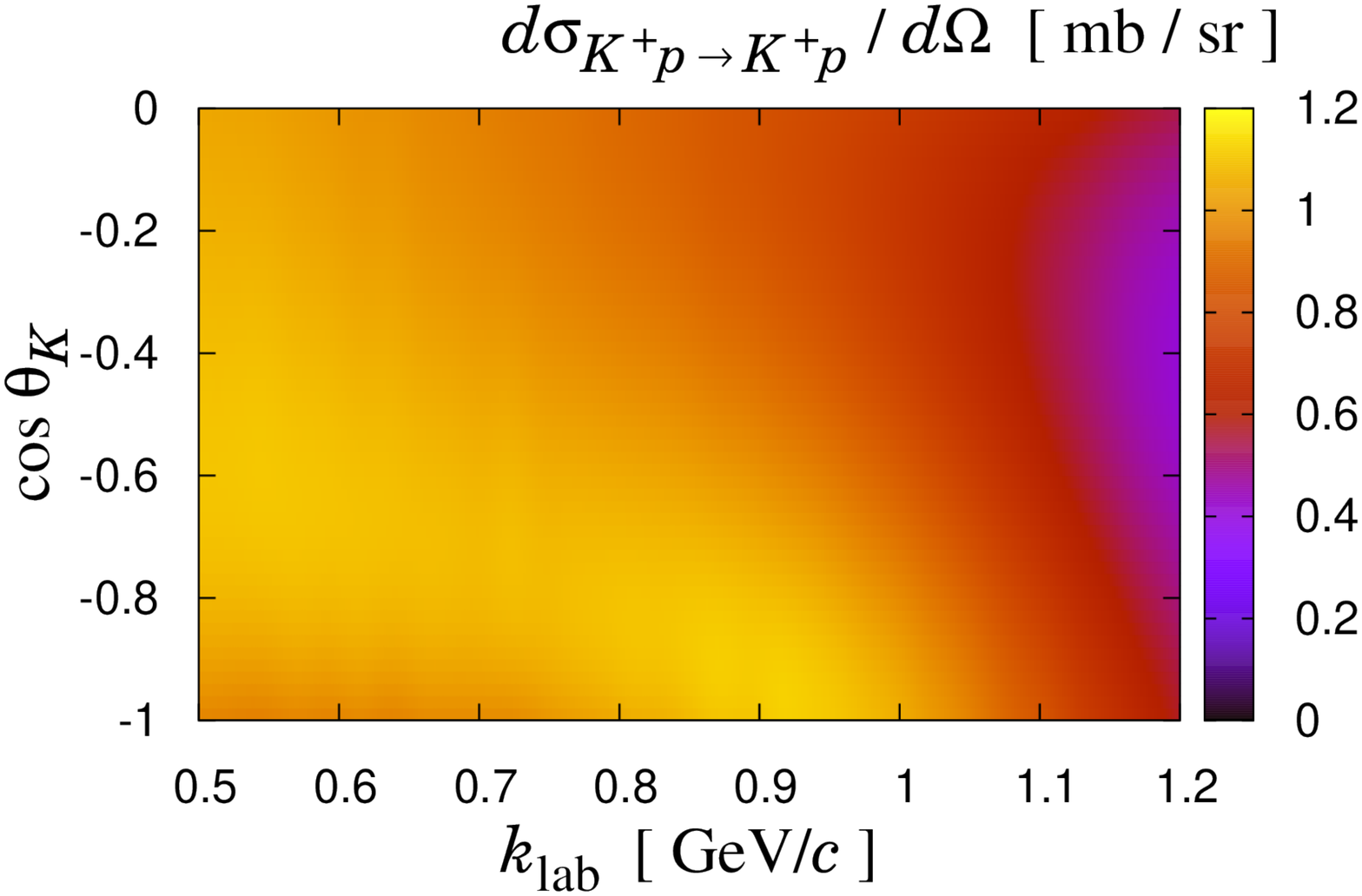}
  \Psfig{7.6cm}{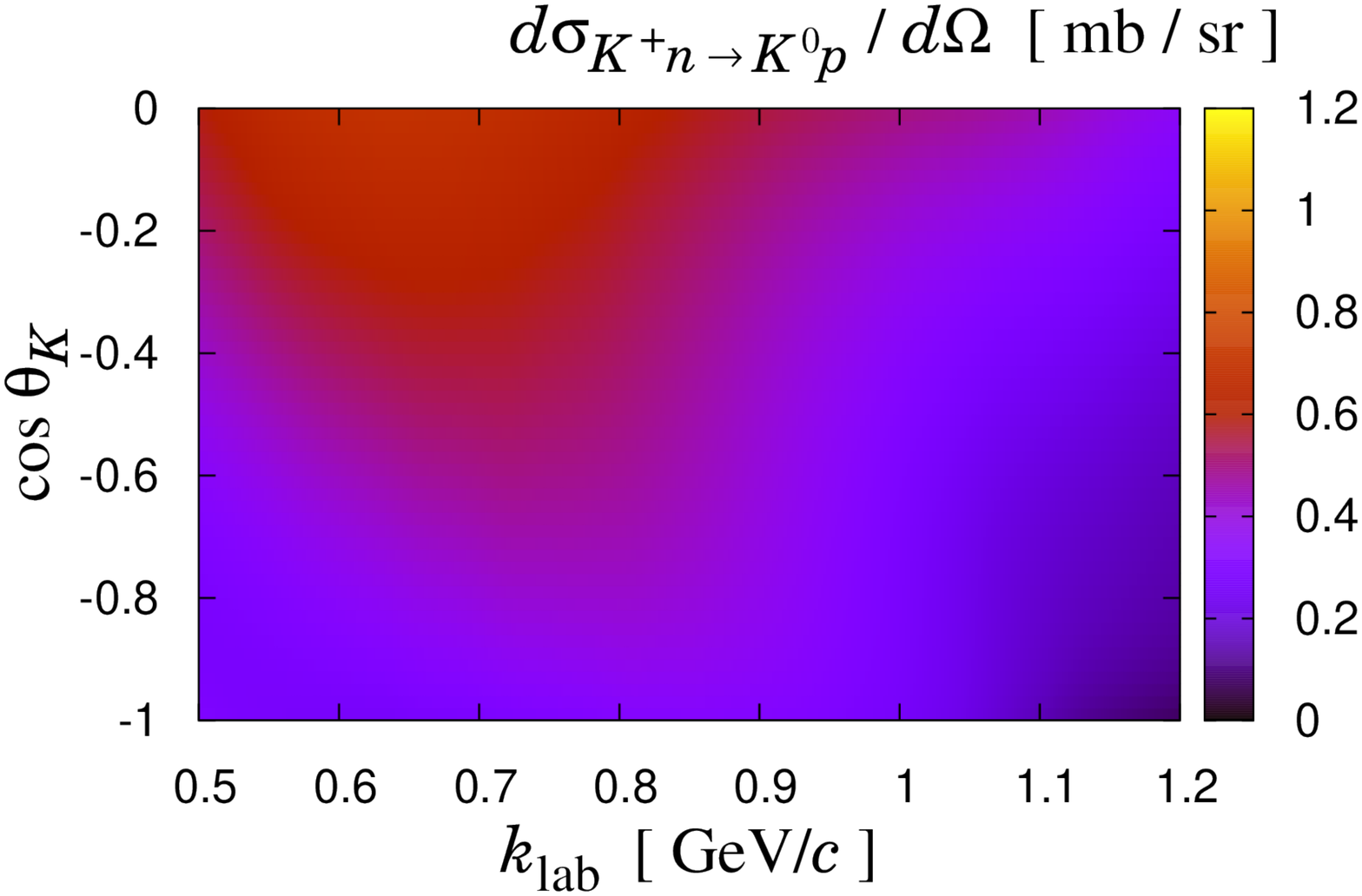}
  \caption{Differential cross sections of the $K^{-} n \to K^{-} n$
    (left top), $K^{-} p \to \bar{K}^{0} n$ (right top), $K^{+} p \to
    K^{+} p$ (left bottom), and $K^{+} n \to K^{0} p$ (right bottom)
    reactions.  The $\bar{K} N \to \bar{K} N$ cross sections are taken
    from a dynamical coupled-channels model in
    Ref.~\cite{Kamano:2014zba}, and the $K N \to K N$ cross sections
    are calculated with the SAID amplitudes~\cite{SAID}.}
  \label{fig:8}
\end{figure}
As shown in the top panels of Fig.~\ref{fig:8}
for the differential cross sections of the $\bar{K} N \to \bar{K} N$
reaction as functions of the initial antikaon momentum $k_{\rm lab}$
and antikaon scattering angle $\theta _{K}$, the $\bar{K} N \to
\bar{K} N$ differential cross section indeed reveals a peak structure
at $k_{\rm lab} \approx 1 \gev / c$ and $\cos _{K} \approx -1$, which
is essential to obtain a slow antikaon with the forward neutron
emission.  

When it comes to the $K N \to K N$ case, the bottom panels of 
Fig.~\ref{fig:8} illustrate the $K^{+} p \to K^{+} p$ and $K^{+} n \to
K^{0} p$ reaction cross sections, which indicates that $k_{\rm lab} =
0.8$--$0.9 \gev /c$ are the best values for the present study of the
$K^{+} d \to K^{0} p p$ reaction.  With these kaon momenta, we obtain
the largest cross sections of the $K^{+} n \to K^{0} p$ and $K^{+} p
\to K^{+} p$ reactions at $\cos \theta _{K} \approx -1$, which
corresponds to the forward proton emission. 
Thus, we fix the initial kaon momentum to be $k_{\rm lab} = 0.85 \gev
/ c$ and compute the $K^{0} p$ invariant mass spectrum of the $K^{+} d
\to K^{0} p p$ reaction.  Note that we take into account here both the
double-step scattering process and the impulse scattering one. 

\begin{figure}[htp]
  \centering
  \Psfig{8.6cm}{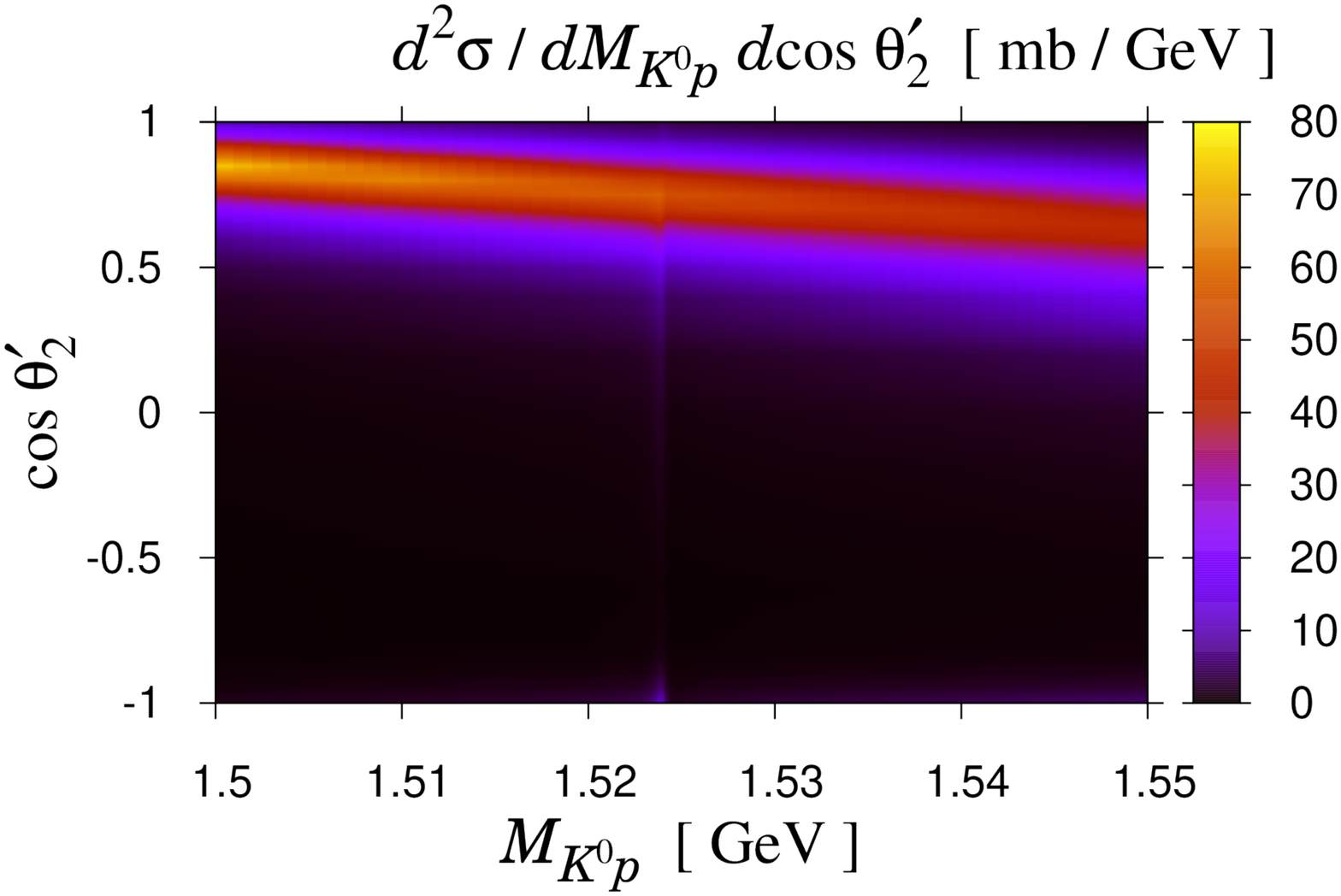}
  \caption{Differential cross section of the $K^{+} d \to K^{0} p p$
    reaction in the impulse plus double scattering processes.  The
    initial kaon momentum is fixed as $k_{\rm lab} = 0.85 \gev /c$.
    The ``$\Theta ^{+}$'' with $J^{P} = 1/2^{+}$ is taken into
    account.}
  \label{fig:9}
\end{figure}

In Fig.~\ref{fig:9}, we draw the results of the differential cross
section in the ``$\Theta ^{+}$'' energy region including the ``$\Theta
^{+}$'' with $J^{P} = 1/2^{+}$.  Figure~\ref{fig:9} exhibits three
structures: a band at $\cos \theta _{2}^{\prime} \gtrsim 0.5$, thin
line at $M_{K^{0} p} = 1.524 \gev$, and sharp peak at $( M_{K^{0} p} ,
\, \cos \theta _{2}^{\prime} ) = ( 1.524 \gev , \, -1 )$.  The first
band structure corresponds to the line in Fig.~\ref{fig:2} and
originates from the impulse scattering contribution. Note, however,
that it was parametrized in terms of the invariant mass of $K^{0}$ and
spectator proton.  The second one, the line structure, represents a
signal of the ``$\Theta ^{+}$''.  The impulse scattering process
cannot generate the line structure because in such a case a highly
off-shell neutron is required.  Therefore, we can conclude that this
line structure is given by the double-step scattering process.  The
third one, which corresponds to the sharp peak around 1.524 GeV and
$\cos \theta _{2}^{\prime} = -1$, arises from the ``$\Theta ^{+}$''
production in the impulse scattering process with a highly off-shell
bound nucleon.  This contribution, however, depends on the tail of the
deuteron wave function in momentum space and thus contains large
theoretical uncertainty.  Therefore, we do not regard this third sharp
peak structure at $( M_{K^{0} p} , \, \cos \theta _{2}^{\prime} ) = (
1.524 \gev , \, -1 )$ as an important one.

\begin{figure}[htp]
  \centering
  \Psfig{10cm}{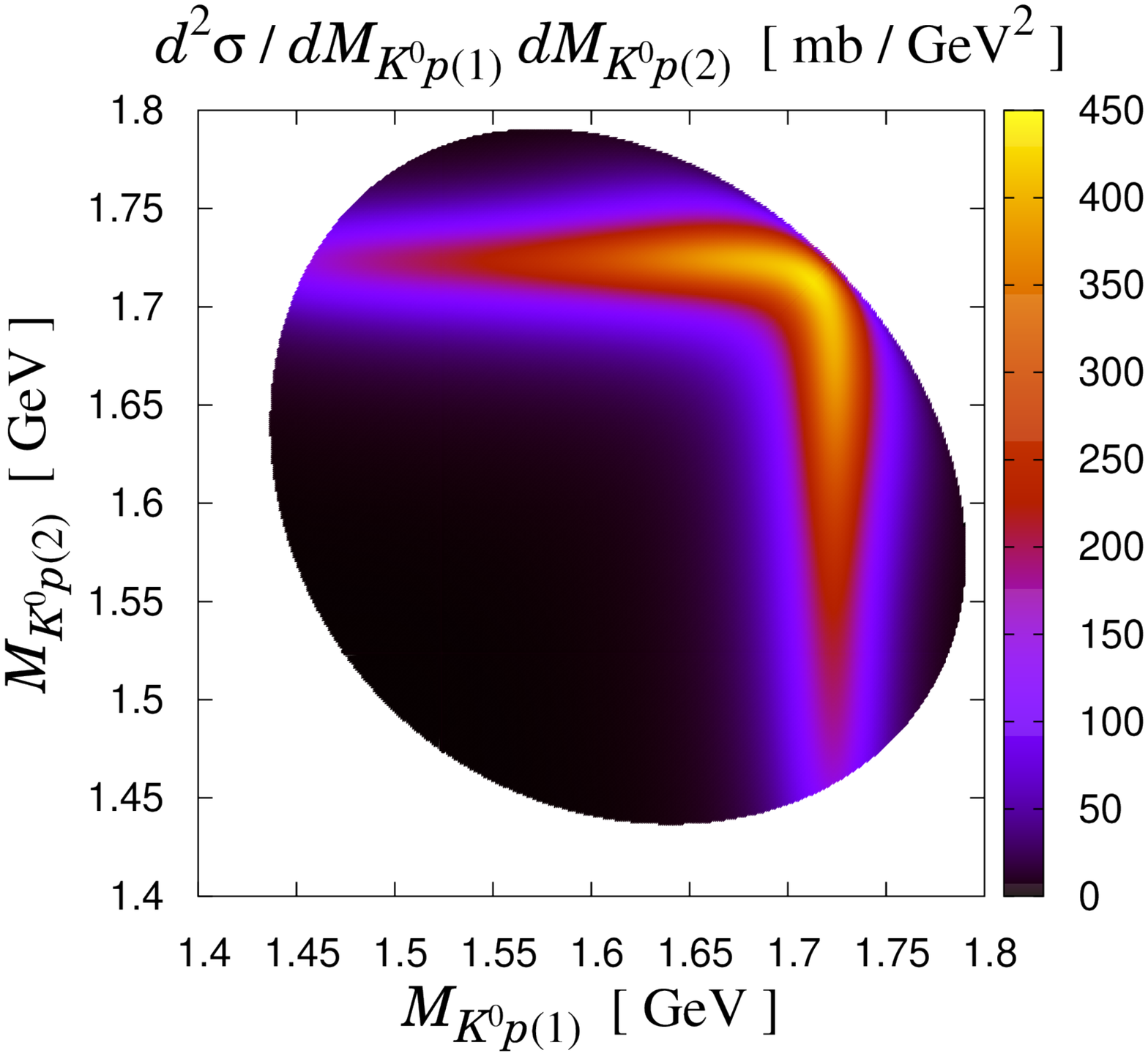}
  \caption{Dalitz plot of the $K^{+} d \to K^{0} p p$ reaction in the
    impulse plus double scattering processes.  The initial kaon
    momentum is fixed as $k_{\rm lab} = 0.85 \gev /c$.  The
    ``$\Theta^{+}$'' with $J^{P} = 1/2^{+}$ is taken into account.}
  \label{fig:10}
\end{figure}

Similarly to the lower kaon momentum case, we can discuss the same
reaction with the Dalitz plot $d^{2} \sigma / d M_{K^{0} p (1)} d
M_{K^{0} p (2)}$.  In Fig.~\ref{fig:10} we show the Dalitz plot with
$k_{\rm lab} = 0.85 \gev / c$ and the ``$\Theta ^{+}$'' spin/parity
$J^{P} = 1/2 ^{+}$.  We can hardly distinguish the ``$\Theta ^{+}$''
signal around $M_{K^{0} p (1)} = 1.524$ in the vertical direction and
$M_{K^{0} p (2)} = 1.524$ in the horizontal direction because the
impulse scattering process dominates the reaction as a whole.
Nevertheless, if we enlarge the Dalitz plot, we can observe the small
sharp structures at $M_{K^{0} p (1)} = 1.524$ in the vertical
direction and $M_{K^{0} p (2)} = 1.524$ in the horizontal direction
as the ``$\Theta ^{+}$'' signal.

\begin{figure}[htp]
  \centering
  \Psfig{8.6cm}{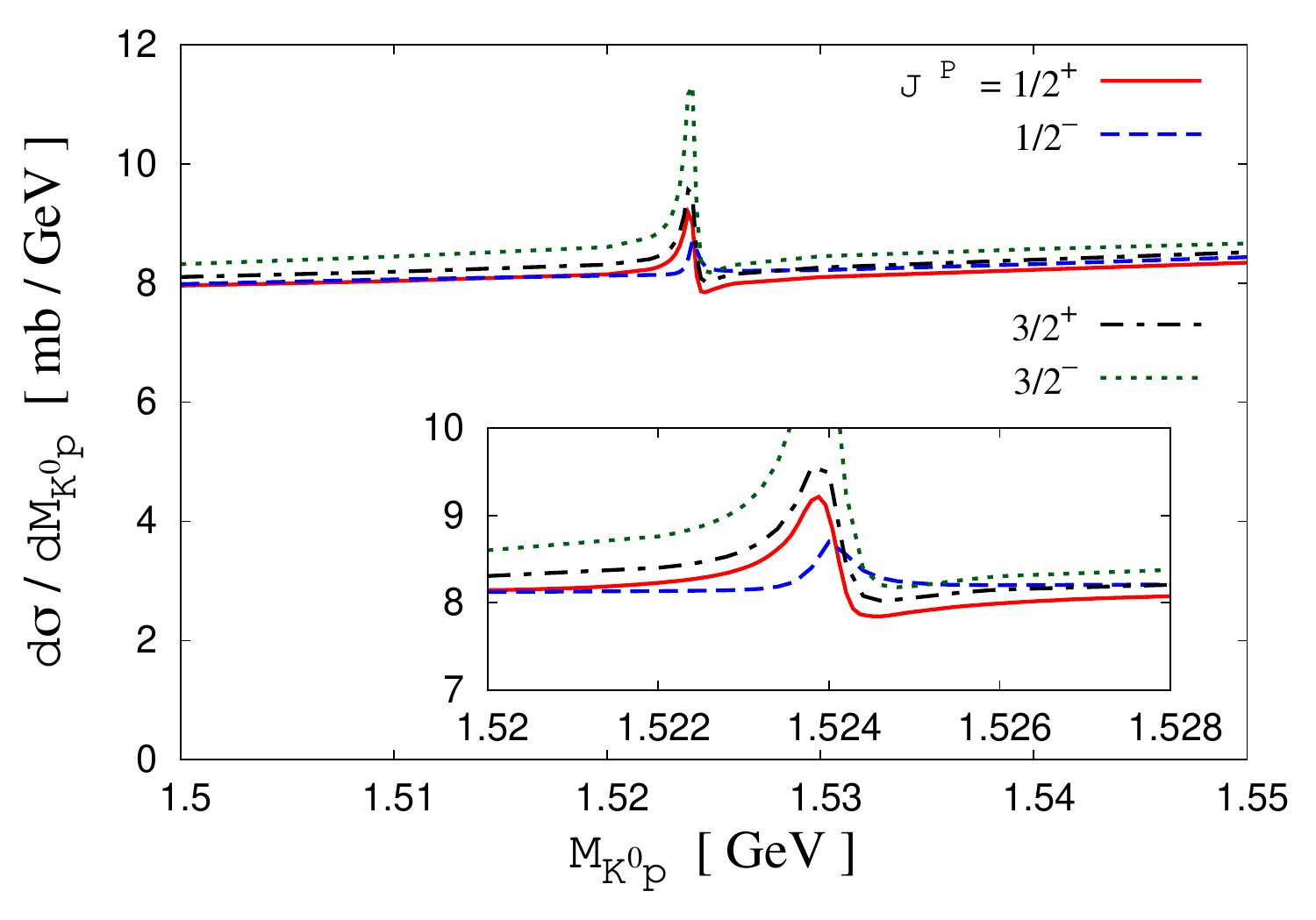}
  \caption{$K^{0} p$ invariant mass spectrum of the $K^{+} d \to K^{0}
    p p$ reaction in the impulse and double-step scattering processes
    with the ``$\Theta ^{+}$'' spin/parity $J^{P} = 1/2^{\pm}$ and
    $3/2^{\pm}$.  The initial kaon momentum is fixed to be $k_{\rm lab} =
    0.85 \gev /c$.  The integral range of the scattering angle is
    given as $0 < \cos \theta _{2}^{\prime} < 1$.  The inset
    represents an enlarged figure.}
  \label{fig:11}
\end{figure}

Now the very important task is to answer how much the signal of the
``$\Theta ^{+}$'' in the double-step scattering process, that is, the
thin line structure in Fig.~\ref{fig:9}, is significant compared to
the impulse scattering contribution parametrized in terms of the
invariant mass of $K^{0}$ and spectator proton, band in
Fig.~\ref{fig:9}. To do that, we integrate the differential cross
section of Fig.~\ref{fig:9} with the integral range $0 < \cos \theta
_{2}^{\prime} < 1$, which results in the $K^{0} p$ invariant mass
spectrum in Fig.~\ref{fig:11}.  Here we introduced the ``$\Theta
^{+}$'' contribution of spin/parity $J^{P} = 1/2^{\pm}$ and
$3/2^{\pm}$.  As shown in Fig.~\ref{fig:11}, with every spin/parity of
the ``$\Theta ^{+}$'', a small ``$\Theta ^{+}$'' signal exists on
  a smooth background.  The smooth background out of the ``$\Theta
  ^{+}$'' energy comes from the broad band at $\cos \theta
  _{2}^{\prime} \gtrsim 0.5$ in Fig.~\ref{fig:9}, so we can see that
  the background originates from the impulse scattering contribution.
  This dominates the cross section (i.e., the integral of the mass
  spectrum) of the reaction.  Nevertheless, a small ``$\Theta ^{+}$''
  signal is not invisible owing to the double-step
processes.  By calculating the excess area of the spectrum in
Fig.~\ref{fig:11} on top of the background from the impulse scattering
contribution, we find that the ``$\Theta^{+}$'' production cross
section turns out to be about $0.9 \microb$, 
$0.4 \microb$, $1 \microb$, and $3 \microb$ for the spin/parity of the
``$\Theta ^{+}$'' $J^{P} = 1/2^{+}$, $J^{P} = 1/2^{-}$, $J^{P} =
3/2^{+}$, and $J^{P} = 3/2^{-}$, respectively.  Therefore, we expect
that in the $K^{+} d \to K^{0} p p$ reaction with higher kaon momenta
$k_{\rm lab} \approx 0.85 \gev / c$ the measurement of the production
cross section $\lesssim 1 \microb$ is required to save the ``$\Theta
^{+}$'' pentaquark.  One can experimentally judge the existence
  of such a narrow peak in the $K^{0} p$ spectrum with the resolution
  of the $K^{0} p$ invariant mass $\sim 1 \mev$. 

\section{Summary and outlook}
\label{sec:4}
In the present work, we have investigated the $K^+ d \to K^0 pp$
reaction as a feasibility study to suggest the kinematical conditions
for the most probable range of the initial kaon momentum and to judge the
existence of the ``$\Theta^+$'' pentaquark in this reaction. We
consider two different dynamical processes for the $K^+ d \to K^0 pp$
reaction, that is, the single-step or impulse scattering process and
the double-step scattering processes. While the first one was already
considered in a previous study, the latter was ignored. In the present
work, we took into account both the processes and scrutinized the
relevant kinematical conditions to each process and their relevances
in the production of the ``$\Theta^+$'' pentaquark. We showed
explicitly that, to produce the ``$\Theta ^{+}$'', the impulse
scattering process is dominant over the double-step scattering process
in lower momentum regions ($k_{\mathrm{lab}}\approx 0.40 \gev /c$),
whereas the double-step one takes over the impulse one in higher
momentum regions ($k_{\mathrm{lab}}\approx 0.85 \gev /c$). We found
that the strength of the bump structure corresponding to the
``$\Theta^+$'' is about a few hundred $\mu\mathrm{b}$ to 1 mb in the
lower momentum region, while it is about 1 $\mu\mathrm{b}$ in the
higher momentum region.

The $K^+$ beam has a unique feature in investigating the existence of
the ``$\Theta^+$'', compared with almost all experiments done
previously. This tentative pentaquark state ``$\Theta^+$'' is strongly
coupled to either $K^+ n$ or $K^0 p$. It implies that the charged
$K^+$ beam provides a chance to produce the ``$\Theta^+$'' by diract
formation. Thus, it is not required to resort to any complicated
methods of experimental analyses to observe the ``$\Theta^+$'', if it
exists. In this sense, the J-PARC is the best place to perform the
ultimate experiments with the $K^+$ beam to put a final period to the
matter of the ``$\Theta^+$'' existence. It is physically worthwhile to
carry out such experiments in the future. If the experiments at the
J-PARC find that ``$\Theta^+$'' does not exist, it will bring any
debate on the existence of the ``$\Theta^+$'' to an end. However, if
the experiments yield any evidence for its existence, it will reignite
interest in the physics of the light pentaquarks.

\section*{Acknowledgments}
The authors want to express their gratitude to M.~Oka and K.~Tanida
for useful discussions. H.-Ch.K.\ is also grateful to M.~V.~Polyakov and
Gh.-S.~Yang for valuable discussions and comments. He is also very
thankful to the members of the
Advanced Science Research Center (ASRC),
Japan Atomic Energy Agency (JAEA) for the hospitality and
supports, where part of the work was carried out. 
H.-Ch.K.\ is supported by Basic Science Research Program
through the National Research Foundation of Korea funded by the
Ministry of Education, Science and Technology (2018R1A2B2001752  and
2018R1A5A1025563 (H.-Ch.K.)).     
A.H.\ is supported in part by JSPS KAKENHI No.~JP17K05441 (C) and
Grants-in-Aid for Scientific Research on Innovative Areas (No.~18H05407).

\appendix


\begin{thebibliography}{99}
\bibitem{Aaij:2015tga} 
  R.~Aaij {\it et al.} [LHCb Collaboration],
  Phys.\ Rev.\ Lett.\  {\bf 115}, 072001 (2015).
  
\bibitem{Aaij:2016phn} 
  R.~Aaij {\it et al.} [LHCb Collaboration],
  Phys.\ Rev.\ Lett.\  {\bf 117}, 082002 (2016).

\bibitem{Aaij:2016ymb} 
  R.~Aaij {\it et al.} [LHCb Collaboration],
  Phys.\ Rev.\ Lett.\  {\bf 117}, 082003 (2016)
  Addendum: [Phys.\ Rev.\ Lett.\  {\bf 117}, 109902 (2016)]
  Addendum: [Phys.\ Rev.\ Lett.\  {\bf 118}, (2017)].

\bibitem{Aaij:2019vzc} 
  R.~Aaij {\it et al.} [LHCb Collaboration],
  Phys.\ Rev.\ Lett.\  {\bf 122}, 222001 (2019).

\bibitem{Aaij:2017nav} 
  R.~Aaij {\it et al.} [LHCb Collaboration],
  Phys.\ Rev.\ Lett.\  {\bf 118}, 182001 (2017).

\bibitem{Yelton:2017qxg} 
  J.~Yelton {\it et al.} [Belle Collaboration],
  Phys.\ Rev.\ D {\bf 97}, 051102 (2018).
  
\bibitem{Kim:2017jpx} 
  H.-Ch.~Kim, M.~V.~Polyakov and M.~Praszalowicz,
  Phys.\ Rev.\ D {\bf 96}, 014009 (2017)
  Addendum: [Phys.\ Rev.\ D {\bf 96}, 039902 (2017)].

\bibitem{Kim:2017khv} 
  H.-Ch.~Kim, M.~V.~Polyakov, M.~Praszalowicz and G.~S.~Yang,
  Phys.\ Rev.\ D {\bf 96}, 094021 (2017)
  Erratum: [Phys.\ Rev.\ D {\bf 97}, 039901 (2018)].

\bibitem{An:2017lwg} 
  C.~S.~An and H.~Chen,
  Phys.\ Rev.\ D {\bf 96}, 034012 (2017).

\bibitem{Wang:2018alb} 
  Z.~G.~Wang and J.~X.~Zhang,
  Eur.\ Phys.\ J.\ C {\bf 78}, 503 (2018).

\bibitem{Diakonov:1997mm} 
  D.~Diakonov, V.~Petrov and M.~V.~Polyakov,
  Z.\ Phys.\ A {\bf 359}, 305 (1997).

\bibitem{Nakano:2003qx} 
  T.~Nakano {\it et al.} [LEPS Collaboration],
  Phys.\ Rev.\ Lett.\  {\bf 91}, 012002 (2003).

\bibitem{Battaglieri:2005er} 
  M.~Battaglieri {\it et al.} [CLAS Collaboration],
  Phys.\ Rev.\ Lett.\  {\bf 96}, 042001 (2006).
  
\bibitem{McKinnon:2006zv} 
  B.~McKinnon {\it et al.} [CLAS Collaboration],
  Phys.\ Rev.\ Lett.\  {\bf 96}, 212001 (2006).

\bibitem{DeVita:2006aaq} 
  R.~De Vita {\it et al.} [CLAS Collaboration],
  Phys.\ Rev.\ D {\bf 74}, 032001 (2006).

\bibitem{Miwa:2006if} 
  K.~Miwa {\it et al.} [KEK-PS E522 Collaboration],
  Phys.\ Lett.\ B {\bf 635}, 72 (2006).

\bibitem{Shirotori:2012ka} 
  K.~Shirotori {\it et al.},
  Phys.\ Rev.\ Lett.\  {\bf 109}, 132002 (2012).

\bibitem{Moritsu:2014bht} 
  M.~Moritsu {\it et al.} [J-PARC E19 Collaboration],
  Phys.\ Rev.\ C {\bf 90}, 035205 (2014).

\bibitem{Shen:2016csu} 
  C.~P.~Shen {\it et al.} [Belle Collaboration],
  Phys.\ Rev.\ D {\bf 93}, 112017 (2016).

\bibitem{Nakano:2008ee} 
  T.~Nakano {\it et al.} [LEPS Collaboration],
  Phys.\ Rev.\ C {\bf 79}, 025210 (2009).

\bibitem{Niiyama:2013dya} 
  M.~Niiyama [LEPS and LEPS II Collaborations],
  Nucl.\ Phys.\ A {\bf 914}, 543 (2013).

\bibitem{Barmin:2013lva} 
  V.~V.~Barmin {\it et al.} [DIANA Collaboration],
  Phys.\ Rev.\ C {\bf 89}, 045204 (2014).

\bibitem{Barmin:2015cta} 
  V.~V.~Barmin {\it et al.} [DIANA Collaboration],
  arXiv:1507.06001 [hep-ex].

\bibitem{Amaryan:2011qc} 
  M.~J.~Amaryan {\it et al.},
  Phys.\ Rev.\ C {\bf 85}, 035209 (2012).

\bibitem{Asratyan:2016qfs} 
  A.~E.~Asratyan and V.~A.~Matveev,
  arXiv:1608.08523 [hep-ex].

\bibitem{Kuznetsov:2006kt} 
  V.~Kuznetsov {\it et al.} [GRAAL Collaboration],
  Phys.\ Lett.\ B {\bf 647}, 23 (2007).
  
\bibitem{Miyahara:2007zz}
  F.~Miyahara {\it et al.},
  Prog.\ Theor.\ Phys.\ Suppl.\  {\bf 168}, 90 (2007).

\bibitem{Kuznetsov:2008hj} 
  V.~Kuznetsov {\it et al.},
  Acta Phys.\ Polon.\ B {\bf 39}, 1949 (2008).

\bibitem{Jaegle:2008ux} 
  I.~Jaegle {\it et al.} [CBELSA and TAPS Collaborations],
  Phys.\ Rev.\ Lett.\  {\bf 100}, 252002 (2008).

\bibitem{Jaegle:2011sw} 
  I.~Jaegle {\it et al.},
  Eur.\ Phys.\ J.\ A {\bf 47}, 89 (2011).

\bibitem{Werthmuller:2013rba} 
  D.~Werthm\"uller {\it et al.} [A2 Collaboration],
  Phys.\ Rev.\ Lett.\  {\bf 111}, 232001 (2013).

\bibitem{Witthauer:2013tkm} 
  L.~Witthauer {\it et al.} [A2 Collaboration],
  Eur.\ Phys.\ J.\ A {\bf 49}, 154 (2013).

\bibitem{Werthmuller:2014thb} 
  D.~Werthm\"uller {\it et al.} [A2 Collaboration],
  Phys.\ Rev.\ C {\bf 90}, 015205 (2014).

\bibitem{McNicoll:2010qk} 
  E.~F.~McNicoll {\it et al.} [Crystal Ball at MAMI Collaboration],
  Phys.\ Rev.\ C {\bf 82}, 035208 (2010)
  Erratum: [Phys.\ Rev.\ C {\bf 84}, 029901 (2011)].

\bibitem{Witthauer:2017get} 
  L.~Witthauer {\it et al.} [A2 Collaboration],
  Phys.\ Rev.\ Lett.\  {\bf 117}, 132502 (2016).

\bibitem{Metag:2019bay} 
  V.~Metag {\it et al.} [CBELSA and TAPS Collaborations],
  EPJ Web Conf.\  {\bf 199}, 02008 (2019).

\bibitem{Polyakov:2003dx} 
  M.~V.~Polyakov and A.~Rathke,
  Eur.\ Phys.\ J.\ A {\bf 18}, 691 (2003).

\bibitem{Kim:2005gz}
H.-Ch.~Kim, M.~Polyakov, M.~Praszalowicz, G.~S.~Yang and K.~Goeke, 
Phys.\ Rev.\ D \textbf{71}, 094023 (2005).

\bibitem{Yang:2018gju} 
  G.~S.~Yang and H.-Ch.~Kim,
PTEP {\bf 2019}, 093D01 (2019).

\bibitem{MartinezTorres:2010zzb} 
  A.~Martinez Torres and E.~Oset,
  Phys.\ Rev.\ Lett.\  {\bf 105}, 092001 (2010).
  
\bibitem{Torres:2010jh} 
  A.~Martinez Torres and E.~Oset,
  Phys.\ Rev.\ C {\bf 81}, 055202 (2010).

\bibitem{Barmin:2003vv} 
  V.~V.~Barmin {\it et al.} [DIANA Collaboration],
  Phys.\ Atom.\ Nucl.\  {\bf 66}, 1715 (2003)
  [Yad.\ Fiz.\  {\bf 66}, 1763 (2003)].

\bibitem{Barmin:2006we} 
  V.~V.~Barmin {\it et al.} [DIANA Collaboration],
  Phys.\ Atom.\ Nucl.\  {\bf 70}, 35 (2007).

\bibitem{Barmin:2009cz} 
  V.~V.~Barmin {\it et al.} [DIANA Collaboration],
  Phys.\ Atom.\ Nucl.\  {\bf 73}, 1168 (2010).

\bibitem{Sibirtsev:2004cf} 
  A.~Sibirtsev, J.~Haidenbauer, S.~Krewald and U.~G.~Meissner,
  Eur.\ Phys.\ J.\ A {\bf 23}, 491 (2005).

\bibitem{Gal:2005cz} 
  A.~Gal and E.~Friedman,
  Phys.\ Rev.\ C {\bf 73}, 015208 (2006).
 
\bibitem{Sibirtsev:2004bg} 
  A.~Sibirtsev, J.~Haidenbauer, S.~Krewald and U.~G.~Meissner,
  Phys.\ Lett.\ B {\bf 599}, 230 (2004).

\bibitem{E949:2014xx} 
  https://www.phy.bnl.gov/e949/analysis/pentaquark/

\bibitem{P09-LoI:2014xx} 
  T.~Nakano {\it et al.}, Letter of Intent for Study of Exotic Hadrons with $S=+1$ and Rare Decay $K^{+} \to \pi ^{+} \nu \bar{\nu}$ with Low-momentum Kaon Beam at J-PARC (the 1st PAC meeting at J-PARC, 2006). \\
  http://j-parc.jp/researcher/Hadron/en/pac\_0606/pdf/p09-Nakano.pdf

\bibitem{JPARC-LoI:2018xx} 
  K.~Tanida {\it et al.}, Letter of Intent for Search for $\Theta ^{+}$
  hypernuclei using $(K^{+}, p)$ reaction (the 4th PAC meeting at J-PARC, 2007).\\
  http://j-parc.jp/researcher/Hadron/en/pac\_0801/pdf/LOI\_Tanida\_pentahyper.pdf

  
\bibitem{Jido:2009jf} 
  D.~Jido, E.~Oset and T.~Sekihara,
  Eur.\ Phys.\ J.\ A {\bf 42}, 257 (2009).
  
\bibitem{Jido:2010rx} 
  D.~Jido, E.~Oset and T.~Sekihara,
  Eur.\ Phys.\ J.\ A {\bf 47}, 42 (2011).
  
\bibitem{Jido:2012cy} 
  D.~Jido, E.~Oset and T.~Sekihara,
  Eur.\ Phys.\ J.\ A {\bf 49}, 95 (2013).

\bibitem{YamagataSekihara:2012yv} 
  J.~Yamagata-Sekihara, T.~Sekihara and D.~Jido,
  PTEP {\bf 2013}, 043D02 (2013).

\bibitem{Gibbs:2004ji} 
  W.~R.~Gibbs,
  Phys.\ Rev.\ C {\bf 70}, 045208 (2004).

\bibitem{Lacombe:1981eg} 
  M.~Lacombe, B.~Loiseau, R.~Vinh Mau, J.~Cote, P.~Pires and R.~de Tourreil,
  Phys.\ Lett.\  {\bf 101B}, 139 (1981).

\bibitem{Machleidt:2000ge} 
  R.~Machleidt,
  Phys.\ Rev.\ C {\bf 63}, 024001 (2001).

\bibitem{Kamano:2016djv} 
  H.~Kamano and T.-S.~H.~Lee,
  Phys.\ Rev.\ C {\bf 94}, 065205 (2016).
  
\bibitem{Yang:2013tka} 
  G.~S.~Yang and H.-Ch.~Kim,
  PTEP {\bf 2013}, 013D01 (2013).

\bibitem{SAID}
  INS Data Analysis Center. INS DAC Services [SAID Program]
  (Institute for Nuclear Studies, Ashburn,
  VA): http://gwdac.phys.gwu.edu.
  
\bibitem{Slater:1961zz} 
  W.~Slater, D.~H.~Stork, H.~K.~Ticho, W.~Lee, W.~Chinowsky, G.~Goldhaber, S.~Goldhaber and T.~O'Halloran,
  Phys.\ Rev.\ Lett.\  {\bf 7}, 378 (1961).

\bibitem{Giacomelli:1972vb} 
  G.~Giacomelli {\it et al.},
  Nucl.\ Phys.\ B {\bf 37}, 577 (1972).

\bibitem{Damerell:1975kw} 
  C.~J.~S.~Damerell {\it et al.},
  Nucl.\ Phys.\ B {\bf 94}, 374 (1975).

\bibitem{Glasser:1977xs} 
  R.~G.~Glasser, G.~A.~Snow, D.~Trevvett, R.~A.~Burnstein, C.~Fu, R.~Petri, G.~Rosenblatt and H.~A.~Rubin,
  Phys.\ Rev.\ D {\bf 15}, 1200 (1977).

\bibitem{Sada:2016nkb} 
  Y.~Sada {\it et al.} [J-PARC E15 Collaboration],
  PTEP {\bf 2016}, 051D01 (2016).

\bibitem{Ajimura:2018iyx} 
  S.~Ajimura {\it et al.} [J-PAC E15 Collaboration],
  Phys.\ Lett.\ B {\bf 789}, 620 (2019).

\bibitem{Sekihara:2016vyd} 
  T.~Sekihara, E.~Oset and A.~Ramos,
  PTEP {\bf 2016}, 123D03 (2016).

\bibitem{Kamano:2014zba} 
  H.~Kamano, S.~X.~Nakamura, T.-S.~H.~Lee and T.~Sato,
  Phys.\ Rev.\ C {\bf 90}, 065204 (2014).
  
\end{thebibliography}
\end{document}